\documentclass[
aps,
pra,
twocolumn,
superscriptaddress,
floatfix
]{revtex4-1}
\usepackage[dvipdfmx]{graphicx}
\usepackage{dcolumn}
\usepackage{bm}
\usepackage{ulem}
\usepackage{amsmath}
\usepackage{amssymb}
\usepackage{txfonts}
\usepackage{hyperref}
\usepackage{color} 
\usepackage{xcolor}
\newcommand{\bs}   {\boldsymbol}

\newcommand{\e}{{\rm e}}
\newcommand{\imag}{{\rm i}}

\hypersetup{
  colorlinks=true,
  linkcolor=[rgb]{0.60,0.00,0.00},
  citecolor=[rgb]{0.00,0.00,0.60},
  urlcolor=[rgb]{0.00,0.00,0.60},
  setpagesize=false
}

\begin{document}

\title{
  Symmetry-adapted variational quantum eigensolver 
}

\author{Kazuhiro~Seki}
\affiliation{Computational Quantum Matter Research Team, RIKEN Center for Emergent Matter Science (CEMS), Saitama 351-0198, Japan}

\author{Tomonori~Shirakawa}
\affiliation{Computational Materials Science Research Team, RIKEN Center for Computational Science (R-CCS),  Hyogo 650-0047,  Japan}

\author{Seiji~Yunoki}
\affiliation{Computational Quantum Matter Research Team, RIKEN Center for Emergent Matter Science (CEMS), Saitama 351-0198, Japan}
\affiliation{Computational Materials Science Research Team, RIKEN Center for Computational Science (R-CCS),  Hyogo 650-0047,  Japan}
\affiliation{Computational Condensed Matter Physics Laboratory, RIKEN Cluster for Pioneering Research (CPR), Saitama 351-0198, Japan}

\begin{abstract}
  We propose a scheme to restore spatial symmetry of Hamiltonian in the variational-quantum-eigensolver (VQE) 
  algorithm for which the quantum circuit structures used usually break the Hamiltonian symmetry. 
  The symmetry-adapted VQE scheme introduced here simply applies the projection operator, 
  which is Hermitian but not unitary, 
  to restore the spatial symmetry in a desired irreducible representation 
  of the spatial group. The entanglement of a quantum state is still represented in a quantum circuit 
  but the nonunitarity of the projection operator is treated classically as postprocessing in the VQE framework.  
  By numerical simulations 
  for a spin-$1/2$ Heisenberg model
  on a one-dimensional ring, 
  we demonstrate that the symmetry-adapted VQE scheme with a shallower quantum circuit 
  can achieve significant improvement in terms of the fidelity of the ground state and 
  has a great advantage in terms of the ground-state energy with decent accuracy, 
  as compared to the non-symmetry-adapted VQE scheme. 
  We also demonstrate that the present scheme can 
  approximate low-lying excited states that can be 
  specified by symmetry sectors, using the same circuit structure for the ground-state calculation. 
\end{abstract}

\date{\today}

\maketitle

\section{Introduction}

Quantum computing has been attracting great interest recently because of 
experimental realizations of quantum devices
~\cite{Nakamura1999,optical_RMP2007,Ladd2010,RevModPhys.85.623,Chow2014,Barends2014,Riste2015,Kelly2015,Arute2019}. 
Simulating quantum many-body systems might be one of the most 
important applications of quantum computing,
due to their potential capability for naturally simulating 
quantum physics and quantum chemistry systems~\cite{Feynman1982}.

A crucial step toward simulating 
quantum many-body systems on quantum computers is
to develop efficient algorithms that might
differ from classical counterparts.
The variational-quantum-eigensolver (VQE) approach~\cite{Peruzzo2014,McClean2016,Kandala2017} 
is likely a promising scheme for simulating 
quantum many-body systems on near-term quantum devices 
including noisy intermediate-scale quantum (NISQ) devices~\cite{Preskill2018}. 
The VQE is a so-called hybrid quantum-classical approach,
where 
the expectation value of a many-body Hamiltonian of interest
with respect to a trial state, represented by a parametrized quantum circuit, is evaluated on quantum computers,
while variational parameters entering in the circuit
are optimized on classical computers by minimizing
the variational energy~\cite{Li2017}.
Here, the number of the variational parameters should be
polynomial in the number of qubits and thus
the optimization on classical computers remains feasible.

Recently, 
quantum algorithms for simulating quantum many-body systems
are vastly proposed, developed, and extended to obtain
not only
ground states~\cite{Mitarai2018PRA,Motta2019,nakanishi2019sequential,parrish2019jacobi}
but also 
excited states~\cite{nakanishi2018subspacesearch,Higgott2019},
excitation spectrum~\cite{Chiesa2019,Parrish2019,kosugi2019construction,calculEndo2019,rungger2019dynamical,
  keen2019quantumclassical,kosugi2019charge},
finite-temperature properties~\cite{Riera2012,Dallaire-Demers2016,zhu2019variational}, and 
non-equilibrium properties~\cite{yoshioka2019variational}. 
A method for simulating 
fermionic particles coupled to bosonic fields
has also been proposed~\cite{Maridin2018PRL,Macridin2018}.
Furthermore, quantum circuits for
preserving symmetry of the Hamiltonian such as  
total spin and time-reversal symmetry~\cite{Sugisaki2016,Sugisaki2019,Liu2019,Gard2019}
have been proposed.
An application of the Grover's search algorithm
for solving a basis-lookup problem of symmetrized
many-body basis states in the exact-diagonalization
method has also been proposed~\cite{Schmitz2020}.  
Moreover, error mitigation schemes have been developed  
for enabling practical applications of the VQE scheme on NISQ devices 
~\cite{BonetMonroig2018,Endo2019mitigation,McArdle2019mitigation}.

In this paper,
we introduce a symmetry-adapted VQE scheme,
which makes use of spatial symmetry of the
Hamiltonian when evaluating the 
expectation value of the Hamiltonian (and also other observables). 
Namely, to symmetrize a quantum state, the standard projection operator~\cite{Inui} is applied 
to a quantum circuit that does not generally preserve the Hamiltonian symmetry. 
The nonunitarity of the projection operator is treated as postprocessing on classical computers in the VQE framework. 
By numerical simulations for a spin-1/2 Heisenberg ring, we show 
that the symmetry-adapted VQE scheme introduced here can better approximate
the ground state with a shallower circuit, 
as compared to the non-symmetrized VQE scheme. 
Moreover, we demonstrate that 
the symmetry-adapted VQE scheme can be used
to approximate low-lying excited states in given symmetry sectors,  
without changing the circuit structure that is 
used for the ground-state calculation.

The rest of the paper is organized as follows. 
In Sec.~\ref{sec.Model},
we define a spin-1/2 Heisenberg model.
In Sec.~\ref{sec.symmetry}, we briefly review
the projection operator and describe
how to implement spatial symmetry operations on
a quantum circuit using SWAP gates. 
In Sec.~\ref{sec.vqe},
we introduce the symmetry-adapted VQE scheme. 
We also describe the natural-gradient-descent (NGD)
method to optimize variational parameters in a quantum circuit subject to
the symmetry projection, which represents a not normalized quantum state. 
In Sec.~\ref{sec.results}, we demonstrate the symmetry-adapted VQE scheme by 
numerical simulations for the spin-1/2 Heisenberg model.   
The paper is summarized 
in Sec.~\ref{sec.conclusions}. 
Appendixes~\ref{AppA} and \ref{AppB} provide details of
a parametrized two-qubit gate and  
a trial wavefunction used in the present VQE simulation, respectively. 
Appendix~\ref{AppC} describes 
that an entangled spin-singlet pair (i.e., one of the Bell states) formed by 
distant qubits can be generated by repeatedly applying a local two-qubit gate 
for finite times. 
Finally, Appendix~\ref{AppD} illustrates
a ground-state-energy evaluation on quantum hardware.  
Throughout the paper, we set $\hbar=1$.

\section{Model}\label{sec.Model}
\begin{center}
  \begin{figure}
    \includegraphics[width=0.5\columnwidth]{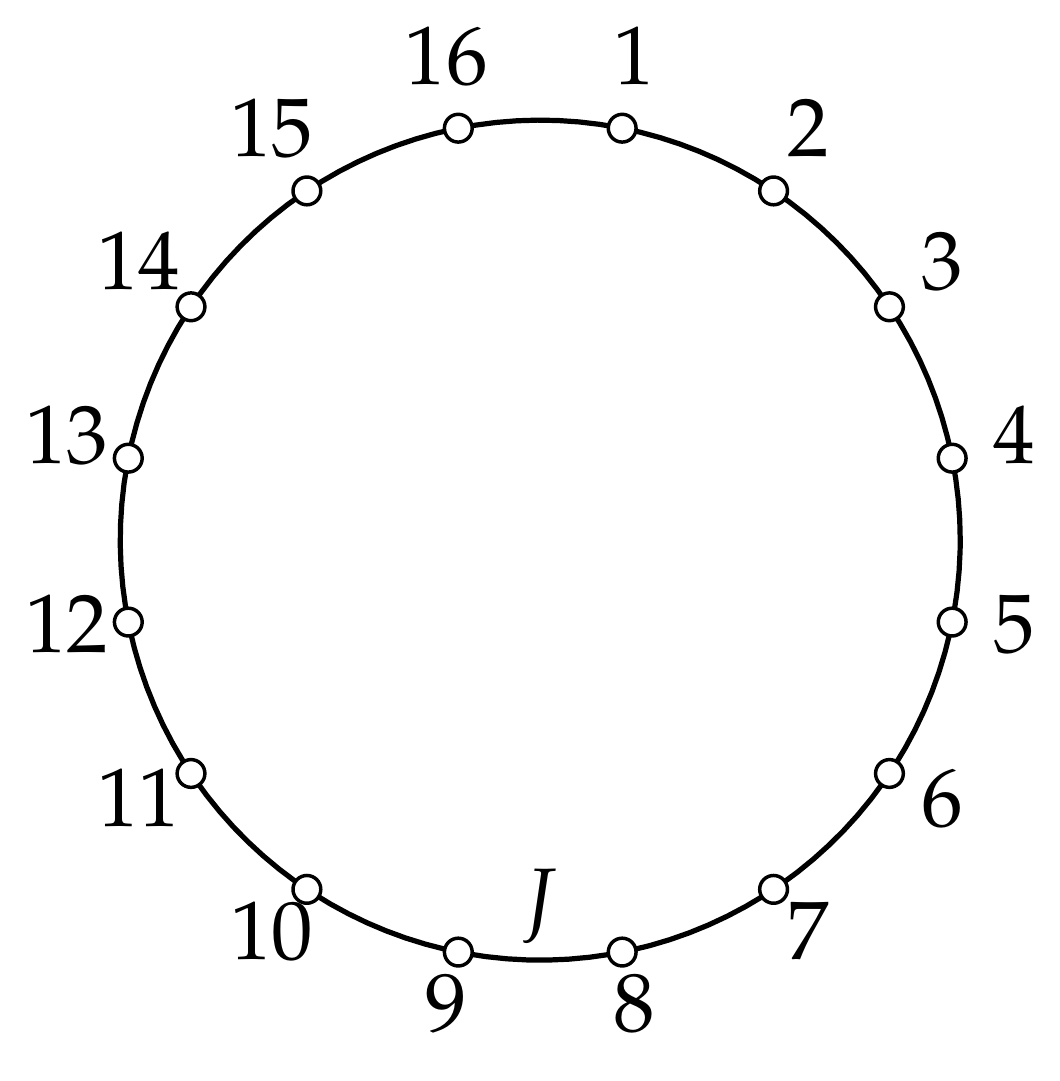}
    \caption{
      Schematic of the
      one-dimensional Heisenberg model      
      with $N=16$ sites
      under the periodic-boundary conditions. 
      The exchange interaction $J$ acts between nearest-neighboring sites at which spin-1/2 spins (i.e., qubits) reside. 
      \label{fig.ring}
    }
  \end{figure}
\end{center}

The Hamiltonian of the spin-1/2 Heisenberg model is given by 
\begin{eqnarray}
  \hat{\mathcal{H}} &=&
  \frac{J}{4}
  \sum_{\langle i,j \rangle} 
  \left(
  \hat{X}_i \hat{X}_j +
  \hat{Y}_i \hat{Y}_j +
  \hat{Z}_i \hat{Z}_j  
  \right) \nonumber \\
  &=&
  \frac{J}{2}
  \sum_{\langle i,j \rangle}
  \left(\hat{\mathcal{P}}_{ij}-\frac{\hat{I}}{2}\right), 
  \label{Ham_SWAP}
\end{eqnarray}
where
$J>0$ is the antiferromagnetic exchange interaction, 
$\langle i,j \rangle$
runs over all nearest-neighbor pairs of qubits $i$ and $j$ 
connected with the exchange interaction $J$, and 
$\hat{X}_{i}$, $\hat{Y}_{i}$, and $\hat{Z}_{i}$
are the Pauli operators acting on the $i$th qubit. 
$\hat{I}$ is the identity operator and  
$\hat{\mathcal{P}}_{ij}$
is the SWAP operator which 
acts on the $i$th and $j$th qubits as 
$\hat{\mathcal{P}}_{ij} |a\rangle_i |b \rangle_j =|b\rangle_i |a \rangle_j $.
The second line in Eq.~(\ref{Ham_SWAP}) follows
from the fact that the inner product of the
Pauli matrices can be written  as
\begin{eqnarray}
  \hat{X}_i \hat{X}_j + \hat{Y}_i \hat{Y}_j + \hat{Z}_i \hat{Z}_j
  =
  \left\{
  \begin{array}{ll}
    \displaystyle 3\hat{I} & (i=j), \\
    \displaystyle 2\hat{\mathcal{P}}_{ij} - \hat{I} & (i\not=j).
    \label{swap}
  \end{array}
  \right.
\end{eqnarray}
Note that $\hat{\mathcal{P}}_{ij}$ is 
Hermitian, unitary, and involutory. 
We consider $\hat{\mathcal{H}}$ on
a one-dimensional periodic chain with $N=16$ sites at which qubits reside (see Fig.~\ref{fig.ring}). 

\section{Spatial symmetries}\label{sec.symmetry}
In this section,
we first briefly review the projection operator that can restore 
the Hamiltonian symmetry of an arbitrary quantum state. 
The projection operator is composed of a set of symmetry operations 
that do not alter the Hamiltonian. 
We then discuss how to implement these symmetry operations 
on a quantum circuit.

\subsection{Projection operator and symmetrized state}\label{sec:projection}
In general, a quantum many-body system possesses its own particular symmetry and 
the Hamiltonian describing such a quantum many-body system is invariant under a set 
of symmetry operations that define the symmetry. These symmetry operations form a 
group, the Hamiltonian symmetry group, and  the symmetry that is relevant to our study here 
is spatial symmetry such as point group symmetry and translational symmetry 
of a lattice where the order of the group is finite. It is well known that an irreducible 
representation of any finite group can be chosen to be unitary~\cite{Inui}.

The projection operator for the $l$th basis
($l=1,\dots,d_\gamma$) of an irreducible representation $\gamma$ 
in a finite group $\mathcal{G}$ is given by  
\begin{equation}
  \hat{P}^{(\gamma)}_l =
  \frac{d_{\gamma}}{|\mathcal{G}|} \sum_{\hat{g}\in\mathcal{G}}
  \left[{\bar D}^{(\gamma)}(\hat{g})\right]^*_{ll} \hat{g},
  \label{eq:pjo}
\end{equation}
where $d_{\gamma}$ is the dimension of the irreducible representation $\gamma$,
$|\mathcal{G}|$ is the order of $\mathcal{G}$,
$\hat{g}$ is a symmetry (unitary) operation in the group $\mathcal{G}$, and
$\left[{\bar D}^{(\gamma)}(\hat{g})\right]_{ll}$ is the $l$th diagonal element of 
a matrix representation for the symmetry operation $\hat{g}$
in the irreducible representation $\gamma$~\cite{Inui,PJO}.
Here, $\hat{g}$ satisfies
$\hat{g} \hat{\mathcal{H}} \hat{g}^{-1} = \hat{\mathcal{H}}$, 
or equivalently 
$[\hat{\mathcal{H}},\hat{g}]=0$.
Thus the projection operator commutes with the Hamiltonian, 
\begin{equation}
  \left[\hat{\mathcal{H}},\hat{P}^{(\gamma)}_l\right]=0.
  \label{eq.Hcommute}
\end{equation}
Note also that the projection operator is
idempotent $(\hat{P}^{(\gamma)}_l)^2 = \hat{P}^{(\gamma)}_l$ and 
Hermitian $(\hat{P}^{(\gamma)}_l)^\dag = \hat{P}^{(\gamma)}_l$,   
but {\it not} unitary.
Eigenvalues of $\hat{P}^{(\gamma)}_l$ are either $0$ or $1$,
implying that it is positive semidefinite. 

For an arbitrary quantum state $|\psi\rangle$, 
the symmetry-projected state
$\hat{P}^{(\gamma)}_l|\psi \rangle$
is indeed the $l$th basis
of the irreducible representation $\gamma$ because, 
for a unitary operator $\hat{g} \in \mathcal{G}$, 
\begin{eqnarray}
\hat g |\psi^{(\gamma)}_l \rangle & = & \hat g \frac{d_{\gamma}}{|\mathcal{G}|} \sum_{\hat{g}'\in\mathcal{G}}
  \left[{\bar D}^{(\gamma)}(\hat{g}')\right]^*_{ll} \hat{g}' |\psi^{(\gamma)}_l \rangle \nonumber \\
  & = & \frac{d_{\gamma}}{|\mathcal{G}|} \sum_{\hat{g}''\in\mathcal{G}}
  \left[{\bar D}^{(\gamma)}(\hat{g}^{-1}\hat{g}'')\right]^*_{ll} \hat{g}''|\psi^{(\gamma)}_l \rangle \nonumber \\
  &=&  \frac{d_{\gamma}}{|\mathcal{G}|} \sum_k \sum_{\hat{g}''\in\mathcal{G}}
  \left[{\bar D}^{(\gamma)}(\hat{g}^{-1})\right]^*_{lk} \left[{\bar D}^{(\gamma)}(\hat{g}'')\right]^*_{kl} \hat{g}'' |\psi^{(\gamma)}_l \rangle \nonumber \\
  &=& \sum_k \left[{\bar D}^{(\gamma)}(\hat{g})\right]_{kl} |\psi^{(\gamma)}_k \rangle,
\end{eqnarray}
where 
\begin{equation}
  |\psi^{(\gamma)}_l \rangle
  = \frac{\hat{P}^{(\gamma)}_l|\psi \rangle}{\sqrt{\langle \psi|\hat{P}^{(\gamma)}_l|\psi \rangle}}.
  \label{symm_norm}
\end{equation}
is the symmetry-projected normalized state, referred to simply as a symmetrized state hereafter, and 
we used $(\hat{P}^{(\gamma)}_l)^2 = \hat{P}^{(\gamma)}_l$ in the first line and 
\begin{equation}
 |\psi^{(\gamma)}_k \rangle = \frac{d_{\gamma}}{|\mathcal{G}|} \sum_{\hat{g}\in\mathcal{G}}
  \left[{\bar D}^{(\gamma)}(\hat{g})\right]^*_{kl} \hat{g} |\psi^{(\gamma)}_l \rangle 
\end{equation}
in the fourth line, which is proved by using the great orthogonality theorem~\cite{Inui}.  

In a one-dimensional representation ($d_\gamma=1$), 
which includes all representations of an Abelian group such as the translation group and 
the identity representation of any point group, 
the projection operator defined in Eq.~(\ref{eq:pjo}) is simply given as 
\begin{equation}
  \hat{P}^{(\gamma)} =
  \frac{1}{|\mathcal{G}|} \sum_{\hat{g}\in\mathcal{G}}
  \chi^{(\gamma)}(\hat{g})^* \hat{g},
\end{equation}
where $\chi^{(\gamma)}(\hat{g})$ is the character 
(i.e., the diagonal element of a matrix representation) 
for the symmetry operation $\hat{g}$
in the irreducible representation $\gamma$ and we omit the subscript ``$l$" in $\hat{P}^{(\gamma)}_l$. 
In this case, the symmetry-projected state
$\hat{P}^{(\gamma)}|\psi \rangle$ for an arbitrary quantum state $|\psi\rangle$ 
is an eigenstate of
a unitary operator $\hat{g} \in \mathcal{G}$
with eigenvalue $\chi^{(\gamma)}(\hat{g})$: 
\begin{equation}
\hat{g}
  \left(\hat{P}^{(\gamma)}|\psi \rangle\right)
  = \chi^{(\gamma)}(\hat{g})
  \left(\hat{P}^{(\gamma)}|\psi \rangle\right). 
\end{equation}

\subsection{Examples of symmetry operations on a quantum circuit}
Translational symmetry of a lattice is described by
an appropriate space group $\mathcal{G}$.
A symmetry operation $\hat{g} \in \mathcal{G}$ 
can be expressed as a product of SWAP operations, 
because $\hat{g}$ simply represents  
a permutation of local (one-qubit) states, 
and any permutation can be expressed as
a product of transpositions.

\begin{center}
  \begin{figure}
    \includegraphics[width=1\columnwidth]{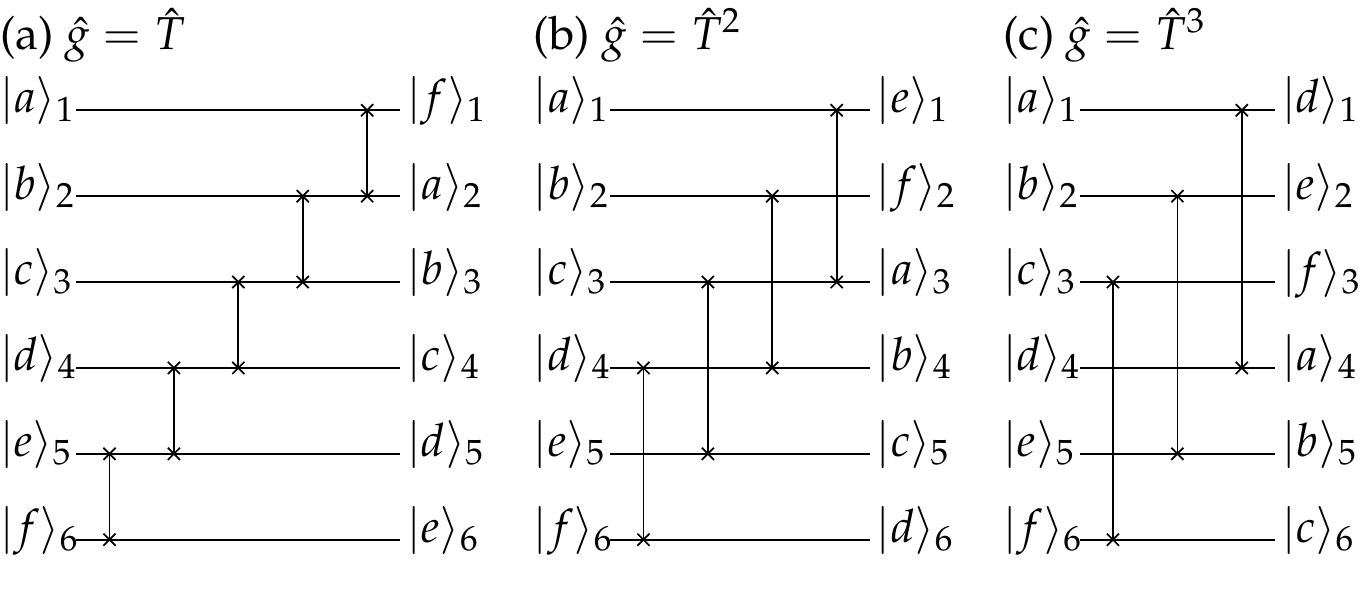}
    \caption{
      Examples of symmetry operations on a six-qubit system
      for 
      (a) one-qubit translation $\hat{T}$,
      (b) two-qubit translation $\hat{T}^2$, and
      (c) three-qubit translation $\hat{T}^3$.
      \label{fig.trns}
    }
  \end{figure}
\end{center}

As examples of $\hat{g}$,
Figs.~\ref{fig.trns}(a), \ref{fig.trns}(b), and \ref{fig.trns}(c) show
translation operations $\hat{T}$, $\hat{T}^2$, and $\hat{T}^3$,
on a six-site ring, respectively. 
Here, $\hat{T}$ is
the one-lattice-space translation
such that $\hat{T}
|a\rangle_1
|b\rangle_2
|c\rangle_3
|d\rangle_4
|e\rangle_5
|f\rangle_6
=
|f\rangle_1
|a\rangle_2
|b\rangle_3
|c\rangle_4
|d\rangle_5
|e\rangle_6
$.
Figure~\ref{fig.trns}(a) shows that
$\hat{T}$ can be expressed as a product of
the SWAP operators as 
$
\hat{T}=
\hat{\mathcal{P}}_{12}
\hat{\mathcal{P}}_{23}
\hat{\mathcal{P}}_{34}
\hat{\mathcal{P}}_{45}
\hat{\mathcal{P}}_{56}
$.
Naively, one can obtain
the one-dimensional $n$-lattice-space translation $\hat{T}^n$
by repeatedly applying the set of the gates 
of the elementary translation $\hat{T}$
for $n$ times ($n$: integer). 
However, 
the representation of a given permutation in terms of a product of transpositions is not unique and 
such a construction of $\hat{T}^n$ may not be optimal
with respect to the number of the SWAP gates. 
The gates shown in Figs.~\ref{fig.trns}(b) and \ref{fig.trns}(c)
are simplified ones for 
$\hat{T}^2$ and $\hat{T}^3$, respectively,
by allowing long-range SWAP gates.
Note that
$\hat{T}^4=(\hat{T}^2)^{-1}$ and 
$\hat{T}^5=\hat{T}^{-1}$
can be obtained by reversing the order of
SWAP operations in Figs.~\ref{fig.trns}(b) and \ref{fig.trns}(a), respectively.

\subsection{General implementation of symmetry operations on a quantum circuit}\label{sec.symop}
\begin{center}
  \begin{figure}
    \includegraphics[width=1\columnwidth]{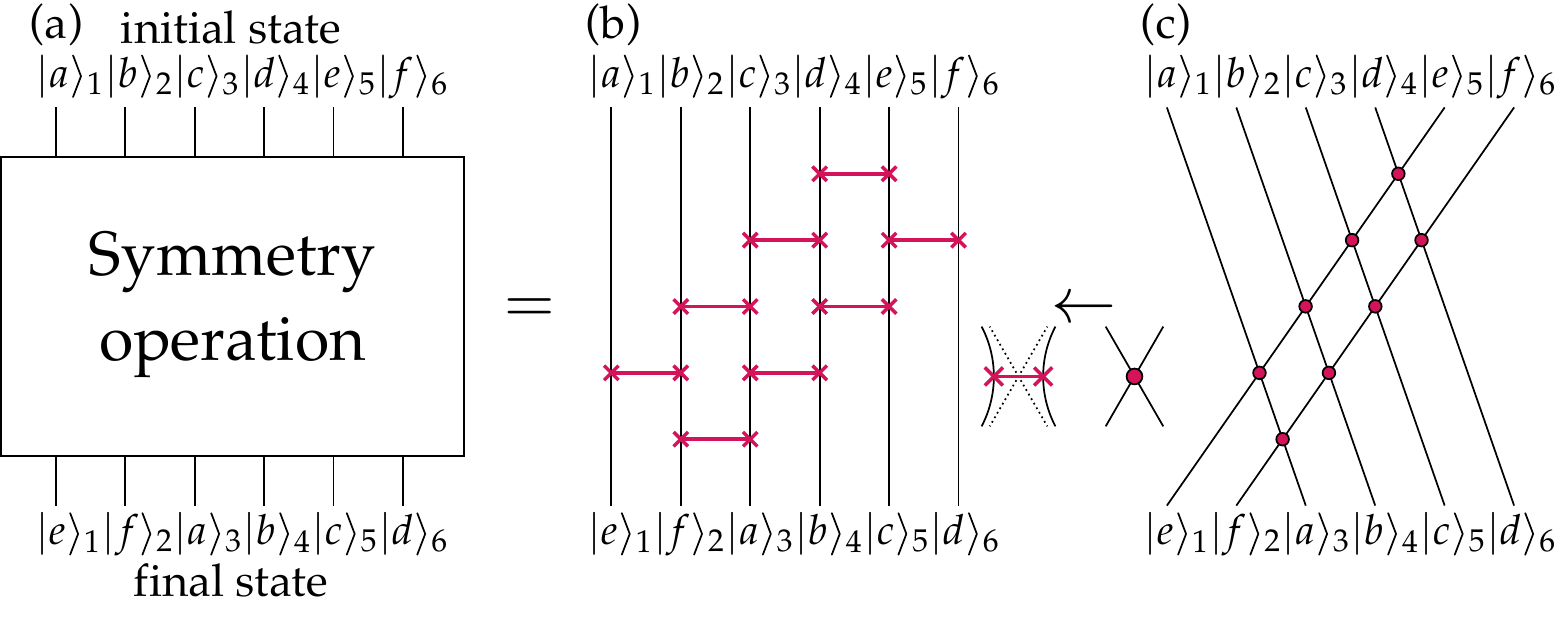}
    \caption{
      An illustration of the ``Amida lottery" construction to implement  
      a general permutation with nearest-neighbor SWAP operations.      
      (a) An oracle of a desired symmetry operation (permutation). 
      (b) A realization of the oracle in (a) with nearest-neighbor SWAP operations.
      The SWAP gates are highlighted with red thick lines. 
      (c) 
      The same one-qubit states in the initial and the final states are connected by the straight line.
      The vertices are highlighted with red circles.
      The vertical lines in panels (a) and (b) represent qubits,
      while the lines in panel (c) are auxiliary. 
      The inset describes how each vertex in panel (c) is replaced with a SWAP gate in panel (b). 
      \label{fig.amida}
    }
  \end{figure}
\end{center}

As a way of implementing generic permutations, 
one can make use of 
the ``Amida lottery'' (sometime also known as ``ghost leg'' or ``ladder climbing'') construction.
Figure~\ref{fig.amida} illustrates 
how to construct a desired permutation
with nearest-neighbor SWAP operations. 
Here, the qubits are depicted as vertical lines   
and the time evolves forward from top to bottom,
to be compatible with the conventional 
two-line notation of a permutation,  
such as, for example,  
\begin{equation}
  \mathcal{S}\equiv
  \left(
  \begin{matrix}
    a & b & c & d & e & f \\
    e & f & a & b & c & d \\
  \end{matrix}
  \right).
\end{equation}

Figure~\ref{fig.amida}(a) is 
an oracle $\hat{g}$ which performs the permutation
$\mathcal{S}$ on one-qubit states, 
\begin{equation}
\hat{g}
|a\rangle_1
|b\rangle_2
|c\rangle_3
|d\rangle_4
|e\rangle_5
|f\rangle_6
=
|e\rangle_1
|f\rangle_2
|a\rangle_3
|b\rangle_4
|c\rangle_5
|d\rangle_6.
\end{equation} 
The oracle can be implemented as a product of
nearest-neighbor-SWAP operations 
shown in Fig.~\ref{fig.amida}(b). 
The circuit structure in Fig.~\ref{fig.amida}(b) can be
obtained with the following procedure
[see Fig.~\ref{fig.amida}(c)]:
(i) draw (unwinding) lines connecting the same one-qubit states 
in the initial and the final states, 
(ii) find all the vertices of the lines drawn,
and
(iii) replace every vertex and its associated four lines, respectively, 
with a SWAP gate and two vertically aligned lines
connected by the SWAP gate 
(see inset of Fig.~\ref{fig.amida}).  

Three remarks are in order. 
First, drawing winding or zigzag lines
in the procedure (i) can produce the same permutation, 
but the resulting circuit may contain unnecessary SWAP operations. 
Second, one can further modify the obtained
circuit structure 
by introducing long-range SWAP gates. 
Third, the inverse permutation, corresponding to $\hat{g}^\dag=\hat{g}^{-1}$,
can be obtained merely by inverting the diagram.

\section{Symmetry-adapted VQE method}\label{sec.vqe}
In this section, 
we first introduce a spin-symmetric quantum state 
that generally breaks spatial symmetry. This is a fundamental step to prepare 
a spin-singlet state. 
Next we describe the symmetry-adapted VQE scheme. 
The procedure is essentially the same as the conventional VQE 
scheme~\cite{Peruzzo2014,McClean2016,Kandala2017}  except that 
the nonunitary projection operator, applied onto a quantum state that is described by a parametrized 
quantum circuit, is treated on classical computers when the variational parameters are updated for the next iteration. 
To optimize the variational parameters, we employ the
NGD method, which requires 
the energy gradient and the metric tensor. We derive these quantities analytically 
for a symmetrized variational quantum state by taking into account 
the fact that the symmetrized state is not normalized 
because the projection operator is not unitary. 
Once the variational parameters in the parametrized quantum circuit are optimized, 
the expectation values of quantities, including those other than the Hamiltonian, for the symmetrized state  
can be evaluated using the resulting circuit by treating the nonunitary projection operator on classical computers as postprocessing.

\subsection{Spin-symmetric trial state}
The total-spin squared operator ${\hat{\bs{S}}}^2$ and
the total magnetization operator $\hat{S}_z$ 
are defined, respectively, as 
$
  \hat{\bs{S}}^2 = \frac{1}{4}
  \sum_{i=1}^{N}
  \sum_{j=1}^{N}
  \left(
  \hat{X}_{i} \hat{X}_{j} +
  \hat{Y}_{i} \hat{Y}_{j} +
  \hat{Z}_{i} \hat{Z}_{j} \right)
  $
and 
$\hat{S}_z = \frac{1}{2}\sum_{i=1}^{N} \hat{Z}_{i}$.
Since 
$[\hat{\mathcal{H}},\hat{\bs{S}}^2]=0$ and 
$[\hat{\mathcal{H}},\hat{S}_z]=0$, 
any eigenstate $|\Psi_n\rangle$ of $\hat{\mathcal{H}}$ is
a simultaneous eigenstate of $\hat{\bs{S}}^2$ and $\hat{S}_z$,
i.e.,
\begin{eqnarray}
\hat{\mathcal{H}} |\Psi_n \rangle &=& E_n    |\Psi_n\rangle, \\
\hat{\bs{S}}^2    |\Psi_n \rangle &=& S(S+1) |\Psi_n\rangle, \\ 
\hat{S}_z         |\Psi_n \rangle &=& S_z    |\Psi_n\rangle, 
\end{eqnarray}
where
$n\, (=0,\ldots,2^{N}-1)$
labels the eigenstates of $\hat{\mathcal{H}}$, and
$E_n$, $S(S+1)$, and $S_z$ are 
the eigenvalues of
$\hat{\mathcal{H}}$, $\hat{\bs{S}}^2$, and $\hat{S}_z$, 
respectively. 
Without loss of generality, we assume that 
$E_0 \leqslant E_1 \leqslant \ldots \leqslant E_{2^{N}-1}$. 
The ground state and the ground-state energy 
of $\hat{\mathcal{H}}$ are thus denoted by $|\Psi_0\rangle$ and $E_0$,
respectively. 

It can be shown that the ground state of the
Heisenberg model is in the subspace of $S=0$~\cite{Lieb1962}. 
To construct a variational state within this subspace,
we first prepare a singlet-pair product state
\begin{equation}
  |\Phi\rangle = \bigotimes_{i=1}^{N/2} |s_{2i-1,2i} \rangle,
  \label{eq.sprod}
\end{equation}
where
$|s_{i,j}\rangle =(|0\rangle_i |1\rangle_j-|1\rangle_i |0\rangle_j)/\sqrt{2}$
is the spin-singlet state (i.e., one of the Bell states) 
formed between the $i$th and $j$th qubits, and therefore 
$|\Phi\rangle$ is spin singlet. 
Then we apply exponential SWAP (eSWAP) gates~\cite{Loss1998,DiVincenzo2000,Brunner2011,Lloyd2014,Lau2016},
each of which is equivalent to the SWAP$^\alpha$ gate 
up to a two-qubit global phase factor~\cite{Fan2005,Balakrishnan2008}, and preserves the spin SU(2) symmetry~\cite{Gard2019,Liu2019}.
The eSWAP gates are parametrized by
a set of angles $\bs{\theta}$ to evolve the state 
from $|\Phi\rangle$ to (an approximation of)
the true ground state $|\Psi_0 \rangle$,  
while keeping the state in the subspace of
$S=0$ during the evolution.

\begin{center}
  \begin{figure}
    \includegraphics[width=.9\columnwidth]{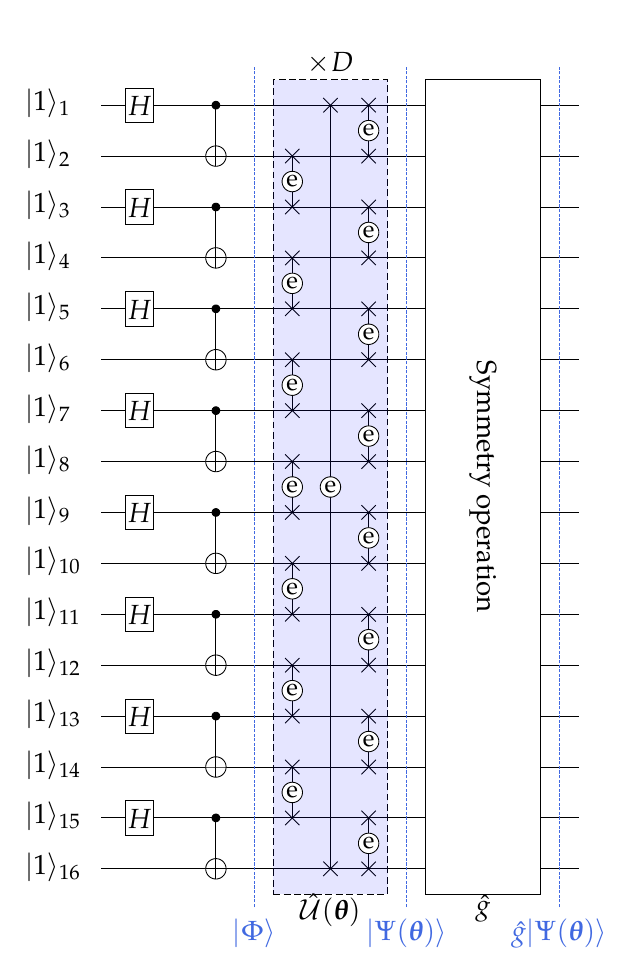}
    \caption{
      The circuit structure that generates a state
      $\hat{g}|\Psi(\bs{\theta})\rangle 
      =\hat{g}\hat{\mathcal{U}}(\bs{\theta})|\Phi\rangle$.
      The symmetry operation $\hat{g}$ can be
      implemented according to
      the scheme described in Sec.~\ref{sec.symop}. 
      The circuit consists of $N$ qubits ($N=16$ in the figure)
      with $D$ layers of gates, each layer being composed of $N$ eSWAP gates (indicated by shaded blue), 
      and the symmetry operation gates. Here, the eSWAP gate is represented by the SWAP gate with symbol ``e''. 
      Since each eSWAP gate contains a single variational parameter, there exist $N\times D$ 
      variational parameters to be optimized in the circuit.  
      \label{fig.circuit}
    }
  \end{figure}
\end{center}

The unitary operator $U_{ij}(\theta)$
corresponding to the eSWAP gate acting on two qubits $i$ and $j$ 
with a parameter $\theta$ is given by
\begin{eqnarray}
  \hat{U}_{ij}(\theta)
  &\equiv&
  \exp(-\imag \theta \hat{\mathcal{P}}_{ij}/2) \nonumber \\
  &=&
  \hat{I} \cos{\frac{\theta}{2}}
  -
  \imag \hat{\mathcal{P}}_{ij} \sin{\frac{\theta}{2}}, 
  \label{eSWAP}
\end{eqnarray}
where the involutority of the SWAP operator 
$\hat{\mathcal{P}}_{ij}^2=\hat{I}$ is used. 
A decomposition of the eSWAP gate in terms of
more elementary gates is described in Appendix~\ref{AppA}. 
By writing the sequence of the eSWAP operations as
\begin{equation}
  \hat{\mathcal{U}}(\bs{\theta}) =
  \prod_{\langle i,j\rangle} \hat{U}_{ij}(\theta_{ij}),
  \label{eq:U}
\end{equation}
with the order of multiplications specified in the circuit construction (see Fig.~\ref{fig.circuit}), 
our trial wavefunction is given by
\begin{equation}
  |\Psi(\bs{\theta}) \rangle = \hat{\mathcal{U}}(\bs{\theta})|\Phi\rangle. 
\end{equation}

Note that $|\Psi(\bs{\theta}) \rangle$ preserves the spin symmetry of the Hamiltonian but not the spatial symmetry, 
as apparently seen in Fig.~\ref{fig.circuit}. 
The order of multiplication of the eSWAP gates in the circuit shown in Fig.~\ref{fig.circuit} is 
motivated by 
an adiabatic evolution of the state from the initial state $|\Phi\rangle$ to the (approximate) ground 
state of $\hat{\cal H}$ in Eq.~(\ref{Ham_SWAP})~\cite{Ho2019}. 
A physical interpretation of the trial wavefunction $|\Psi(\bs{\theta}) \rangle$ 
in conjunction with
a resonating-valence-bond (RVB) state~\cite{Pauling1933,Anderson1973,Fazekas1974},
a superposition of a 
great number of singlet-pair product states~\cite{Oguchi1989},
known as one of the best variational states to describe quantum many-body states~\cite{Becca_Sorella_book},
is discussed in Appendix~\ref{AppB}.
Note that, as shown in Appendix~\ref{AppC}, a 
spin-singlet pair formed by qubits that are separated even at the largest distance
can be generated in $|\Psi(\bs{\theta})\rangle$ 
with $D \sim N/4$, where $D$ is the number of layers, each layer being composed of 
$N$ eSWAP gates (see Fig.~\ref{fig.circuit}).

\subsection{Energy expectation value}
Although $|\Psi(\bs{\theta}) \rangle$ is symmetric in the spin space,
generally it breaks the spatial symmetry of Hamiltonian because of a 
particular structure of the circuit. 
As described in Sec.~\ref{sec:projection}, 
we apply the projection operator ${\hat P}^{(\gamma)}$ to symmetrize $|\Psi(\bs{\theta}) \rangle$~\cite{note1}. 
The resulting symmetrized variational state with the irreducible representation $\gamma$ is 
\begin{equation}
  |\Psi^{(\gamma)}(\bs{\theta}) \rangle
  = 
  \frac{\hat{P}^{(\gamma)}}
       {\sqrt{\mathcal{N}(\bs{\theta})}}
       |\Psi(\bs{\theta}) \rangle,
       \label{eq.symm_var}
\end{equation}
where 
\begin{equation}
  \mathcal{N}(\bs{\theta})
  =\langle \Psi(\bs{\theta})|\hat{P}^{(\gamma)}|\Psi(\bs{\theta}) \rangle. 
\end{equation}
Note that $\mathcal{N}(\bs{\theta}) \geqslant 0$ because
the projection operator 
$\hat{P}^{(\gamma)}$ is a positive semidefinite operator.
The corresponding variational energy is given by 
\begin{eqnarray}
  E^{(\gamma)}(\bs{\theta}) &\equiv&
  E[\Psi^{(\gamma)}(\bs{\theta})]  \notag \\
  &\equiv&
  \langle
  \Psi ^{(\gamma)}(\bs{\theta})
  |
  \hat{\mathcal{H}}
  |
  \Psi ^{(\gamma)}(\bs{\theta})
  \rangle
  \notag \\
  &=&
  \frac
      {\langle \Psi(\bs{\theta})| \hat{\mathcal{H}}\hat{P}^{(\gamma)}|\Psi(\bs{\theta}) \rangle}
      {\langle \Psi(\bs{\theta})| \hat{P}^{(\gamma)}|\Psi(\bs{\theta}) \rangle}
      \notag \\
      &=&
      \frac
      {\sum_{\hat{g}\in \mathcal{G}} \chi^{(\gamma)}(\hat{g})^* \langle \Psi(\bs{\theta})| \hat{\mathcal{H}}\hat{g}|\Psi(\bs{\theta}) \rangle}
      {\sum_{\hat{g}\in \mathcal{G}} \chi^{(\gamma)}(\hat{g})^* \langle \Psi(\bs{\theta})| \hat{g}|\Psi(\bs{\theta}) \rangle}. 
      \label{eq.Energy}
\end{eqnarray}
In the symmetry-adapted VQE scheme, the matrix elements in
the numerator and the denominator in Eq.~(\ref{eq.Energy})
are evaluated on quantum computers 
by, for example, introducing one ancilla qubit~\cite{Li2017PRX,Romero2018,Mitarai2019,Dallaire-Demers2019}. 
This can be done efficiently because
$\hat{\mathcal{H}}$ is a sum of unitary operators and
$\hat{g}$ is a unitary operator as well. The sum over the group operations $\hat g$, 
the order of $\mathcal{G}$ being $\mathcal{O}(N)$, is performed on classical computers as postprocessing~\cite{AV}. 

It should be noted that 
the linear combination of unitary operators can also be implemented
with a circuit described in, e.g., Ref.~\cite{Childs2012}.
The advantage of such a circuit is that it can 
generate the symmetrized state $\hat{P}^{(\gamma)}|\Psi(\bs{\theta})\rangle$ directly 
without introducing the postprocessing. 
However, one major disadvantage of such a circuit, particularly in the current NISQ era, 
is that the circuit structure becomes much more complicated than the one proposed here because 
it requires $\log_2 |\mathcal{G}|$ ancilla qubits and 
$|\mathcal{G}|$ controlled-unitary operations, in addition to the gates necessary 
to describe $|\Psi(\bs{\theta})\rangle$ shown in Fig.~\ref{fig.circuit}.

\subsection{Natural-gradient-descent optimization}
The variational parameters $\bs{\theta}$ are optimized
by minimizing $E^{(\gamma)}(\bs{\theta})$ 
with the NGD optimization~\cite{Amari1998}.
Starting from chosen (e.g., random)
initial parameters $\bs{\theta}_1$, 
the NGD optimization at the $k$th iteration updates 
the variational parameters as
\begin{equation}
  \bs{\theta}_{k+1} = \bs{\theta}_k -
  \alpha
  \left[
    {\rm Re}\bs{G}^{(\gamma)}(\bs{\theta}_k)
    \right]^{-1}
  \bs{\nabla} E^{(\gamma)}(\bs{\theta}_k), 
  \label{eq:step}
\end{equation}
where $\alpha$ is a parameter for tuning the step width (i.e., a learning rate) and 
\begin{eqnarray}
  [\bs{G}^{(\gamma)}(\bs{\theta})]_{ij}
  &\equiv&
  \left[\bs{G}[\Psi^{(\gamma)}(\bs{\theta})] \right]_{ij}
  \notag \\
  &=&
  \langle
  \partial_{\theta_i} \Psi^{(\gamma)}(\bs{\theta})|
  \partial_{\theta_j} \Psi^{(\gamma)}(\bs{\theta})
  \rangle
   \notag \\
  &-&
  \langle 
  \partial_{\theta_i} \Psi^{(\gamma)}(\bs{\theta})|
  \Psi^{(\gamma)}(\bs{\theta})
  \rangle
  \langle
  \Psi^{(\gamma)}(\bs{\theta})|
  \partial_{\theta_j} \Psi^{(\gamma)}(\bs{\theta})
  \rangle
  \label{def.metric}
\end{eqnarray}
is the metric tensor~\cite{provost1980}
of the variational-parameter ($\bs{\theta}$) space
associated with the normalized 
state $|\Psi^{(\gamma)}(\bs{\theta})\rangle$.
Since $\bs{G}^{(\gamma)}(\bs{\theta})$
is positive semidefinite, $\alpha$ has to be
chosen positive to minimize the variational energy.
In the numerical simulations shown in Sec.~\ref{sec.results}, we set $\alpha=0.1/J$.

We should note that essentially the same optimization scheme, which
takes into account the geometry of the
wavefunction in the variational parameter space,
has been introduced as the stochastic-reconfiguration
method and applied successfully with the variational
Monte Carlo technique for correlated
electron systems~\cite{Sorella2001,Casula2003,Yunoki2006}.
An equivalence between the stochastic-reconfiguration method and 
the real- and imaginary-time evolution of a variational state has been pointed out~\cite{Carleo2012,Carleo2014,Ido2015,Takai2016,Nomura2017}.
On the other hand, very recently, as a optimization method, the imaginary-time evolution of a variational 
quantum state has been proposed in the context of VQE approach~\cite{endo2018variational,McArdle2019,Jones2019}.
This method was later recognized to be essentially the same as the NGD optimization of a parametrized 
quantum circuit~\cite{stokes2019quantum,yamamoto2019natural}.

\subsection{Energy gradient and metric tensor}\label{dEandG}
The energy gradient 
$\bs{\nabla}E^{(\gamma)}(\bs{\theta})$ in Eq.~(\ref{eq:step})
and the metric tensor
$\bs{G}^{(\gamma)}(\bs{\theta})$ in Eq.~(\ref{def.metric})
are now expressed in terms of the circuit (non-symmetrized) state
$|\Psi(\bs{\theta}) \rangle$ and its derivative
$|\partial_{\theta_i}\Psi(\bs{\theta}) \rangle$. 
For this purpose, first we can readily show that the derivative of the 
symmetrized state, 
$|\partial_{\theta_i}\Psi^{(\gamma)}(\bs{\theta}) \rangle$, 
can be expressed as 
\begin{eqnarray}
  |\partial_{\theta_i} \Psi^{(\gamma)}(\bs{\theta})\rangle
  &=&
  \frac{\hat{P}^{(\gamma)}}
       {\sqrt{\mathcal{N}(\bs{\theta})}}
       \left[
    |\partial_{\theta_i} \Psi(\bs{\theta}) \rangle
    -{\rm Re}\mathcal{A}_i(\bs{\theta})
    |\Psi(\bs{\theta})\rangle
    \right]
       \label{dPsi}
\end{eqnarray}
with 
\begin{eqnarray}
  \mathcal{A}_i(\bs{\theta})
  &=&\frac{\langle \Psi(\bs{\theta})| \hat{P}^{(\gamma)} |
    \partial_{\theta_i} \Psi(\bs{\theta}) \rangle}{\mathcal{N}(\bs{\theta})}.
  \label{Berry}
\end{eqnarray}
Note that the real part of $\mathcal{A}_i(\bs{\theta})$ is related to 
the logarithmic derivative of the norm: 
\begin{equation}
\partial_{\theta_i} \ln \mathcal{N}(\bs{\theta})
=  2{\rm Re} \mathcal{A}_i(\bs{\theta}), 
\end{equation}
and the imaginary part of $\mathcal{A}_i(\bs{\theta})$ 
is related to the Berry connection: 
\begin{equation}
\langle \Psi^{(\gamma)}(\bs{\theta})|
\partial_{\theta_i} \Psi^{(\gamma)}(\bs{\theta})\rangle
= 
\mathcal{A}_i(\bs{\theta}) - {\rm Re}\mathcal{A}_i(\bs{\theta})  
=
\imag  {\rm Im} \mathcal{A}_i(\bs{\theta}). 
\end{equation}
  
From Eq.~(\ref{dPsi}),
the derivative of the variational energy $E^{(\gamma)}(\bs{\theta})$ can be expressed as 
\begin{equation}
  \partial_{\theta_i}  E^{(\gamma)}(\bs{\theta})
    =
    2{\rm Re}
    \left[
      \frac{
        \langle \Psi(\bs{\theta})|
        \hat{P}^{(\gamma)}
        \hat{\mathcal{H}}
        |\partial_{\theta_i} \Psi(\bs{\theta}) \rangle
      }
           {\mathcal{N}(\bs{\theta})}       
           -\mathcal{A}_i(\bs{\theta})E^{(\gamma)}(\bs{\theta})
           \right].
    \label{dE}
\end{equation}
Similarly, by substituting
Eq.~(\ref{dPsi}) into Eq.~(\ref{def.metric}),  
we can show  
that 
the metric tensor $[\bs{G}^{(\gamma)}(\bs{\theta})]_{ij}$ is now given as 
\begin{eqnarray}
  [\bs{G}^{(\gamma)}(\bs{\theta})]_{ij}
    &=&
    \frac{
    \langle
    \partial_{\theta_i} \Psi(\bs{\theta})
    |
    \hat{P}^{(\gamma)}
    |
    \partial_{\theta_j} \Psi(\bs{\theta})
    \rangle
    }
    {\mathcal{N}(\bs{\theta})}
    -
    \mathcal{A}_i^*(\bs{\theta}) \mathcal{A}_j(\bs{\theta}).  
    \label{Metric}
\end{eqnarray}
Note that Eqs.~(\ref{dPsi}), (\ref{Berry}), (\ref{dE}), and (\ref{Metric}) 
are generic form for the state subject to
the symmetry-projection operator. 

For numerical simulations, 
to evaluate the derivatives of the trial state, 
we employ the parameter-shift rule 
for the (non-symmetrized) state  
\begin{equation}
  |\partial_{\theta_i} \Psi(\bs{\theta})\rangle = \frac{1}{2}|
  \Psi(\bs{\theta}+\pi\bs{e}_i)\rangle, 
\end{equation}
which readily follows from Eq.~(\ref{eSWAP}). 
Here, $\bs{e}_i$ is the unit vector
whose $i'$th entry is given by $[\bs{e}_i]_{i'}=\delta_{ii'}$.
We should also note that our numerical simulations in the next section employ the NGD optimization 
because, as described above, this optimization method has been repeatedly proved to be currently the best method 
for optimizing a variational wavefunction with many variational parameters in the variational Monte Carlo technique 
for quantum many-body systems, when up to the first order derivative of the 
variational energy is available~\cite{Becca_Sorella_book}. If we employ this optimization method in the real experiment, 
we have to evaluate, in addition to the matrix elements in the numerator and the denominator in Eq.~(\ref{eq.Energy}), 
several other quantities appearing in Eqs.~(\ref{dE}) and (\ref{Metric}) on quantum computers. 
However, the use of the NGD optimization is not necessarily required in the symmetry-adapted VQE scheme and
we can always adopt a simpler optimization method without even using the first derivative of the variational energy.

\section{Results}\label{sec.results}

Here we demonstrate the symmetry-adapted VQE approach by numerically simulating the spin-1/2 Heisenberg ring. 

\subsection{Ground-state energy}\label{sec:gs}
\begin{center}
  \begin{figure}
    \includegraphics[width=1\columnwidth]{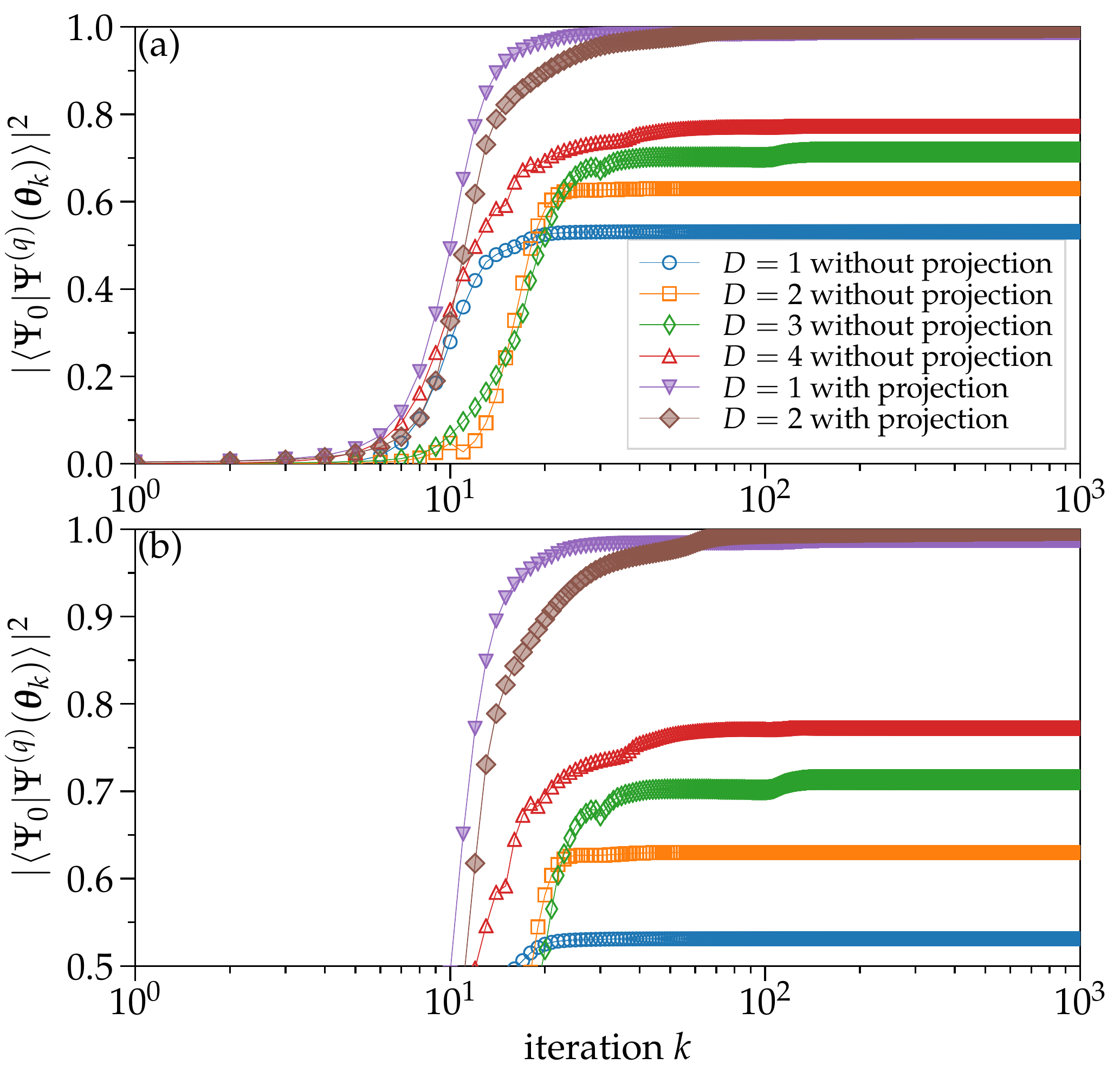}
    \caption{
      Semi-log plot of 
      the fidelity $\left|\langle\Psi_0|\Psi^{(q)}(\bm{\theta}_k)\rangle\right|^2$ of the ground state 
      for the spin-1/2 Heisenberg ring with $N=16$ 
      as a function of the NGD iteration $k$ in Eq.~(\ref{eq:step}). 
      The results with different number of layers ($D$),
      and with (filled symbols) and without (empty symbols)
      use of the translational symmetry, are shown.
      (b) Enlarged figure of panel (a). 
      $|\Psi_0\rangle$ is the exact ground state and $|\Psi^{(q)}(\bm{\theta}_k)\rangle$
      is an approximate ground state obtained 
      after the $k$th iteration of optimizing the variational parameters in the circuit. The number of total variational 
      parameters is $N\times D$. The initial parameters $\bs{\theta}_1$ are set randomly and we use
      the same initial parameters $\bs{\theta}_1$ for all the simulations shown here when $D$ is the same. 
      \label{fig.fidelity}
    }
  \end{figure}
\end{center}

\begin{center}
  \begin{figure}
    \includegraphics[width=1\columnwidth]{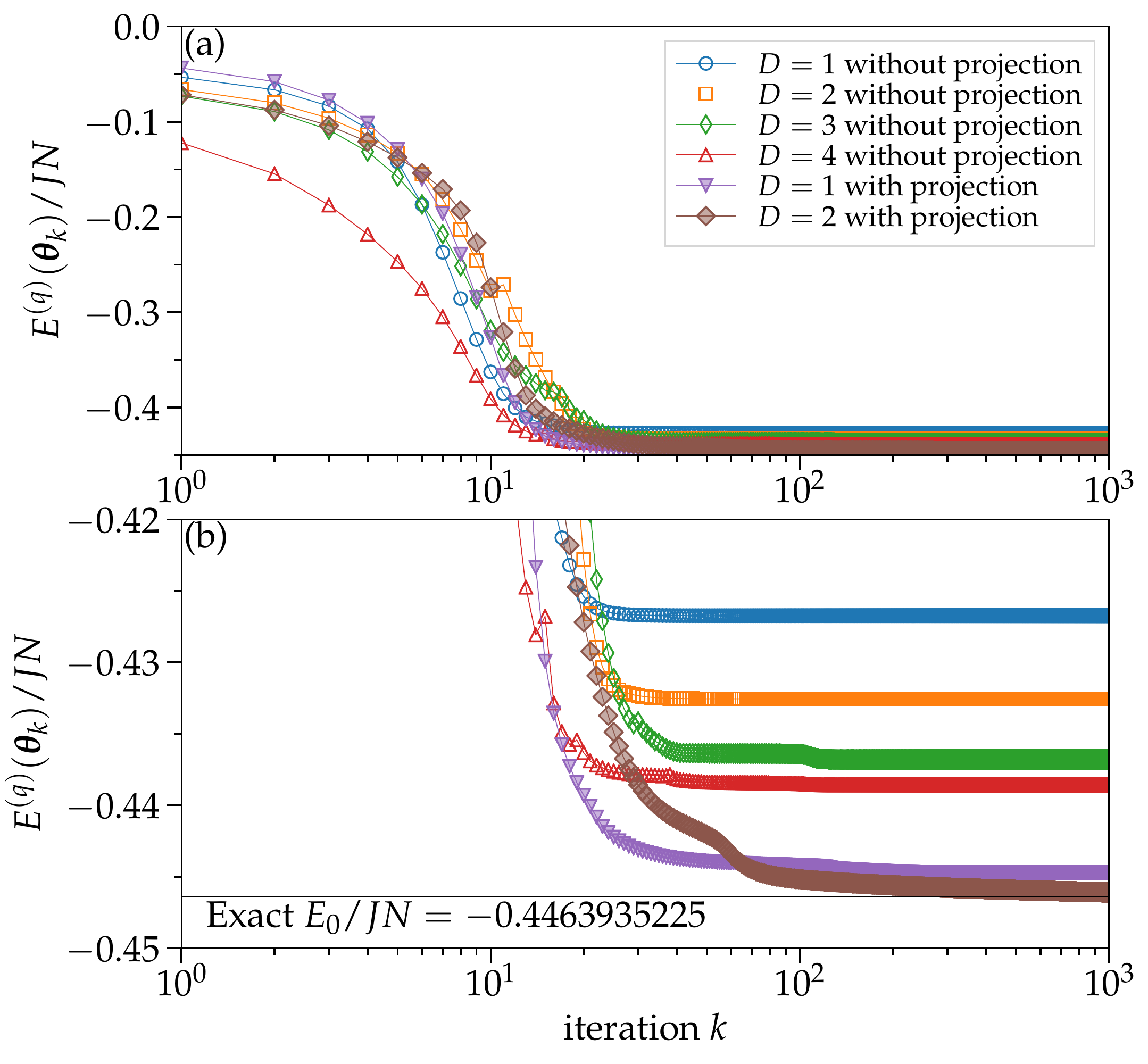}
    \caption{
      Same as Fig.~\ref{fig.fidelity} but for the variational energy of the ground state. 
      The horizontal line in (b) indicates
      the exact ground-state energy $E_0$. 
      \label{fig.energy}
    }
  \end{figure}
\end{center}

Figures~\ref{fig.fidelity} and \ref{fig.energy} show 
a typical behavior of the fidelity and the variational energy $E^{(\gamma)}(\bs{\theta}_k)$, respectively, 
for $N=16$ as a function of the NGD iteration $k$ in Eq.~(\ref{eq:step}). 
Here, we use the translational symmetry of the Hamiltonian that forms 
the cyclic group
$\mathcal{G}=\{\hat{T}^1,\hat{T}^2,\ldots,\hat{T}^N\}$ 
with $|\mathcal{G}|=N$. 
The character associated with the operation $\hat{T}^n$ is given by
\begin{equation}
  \chi^{(q)}(\hat{T}^n)=\e^{\imag q n}, 
\end{equation}
where 
$q=2 \pi m /N$ with $m=-N/2+1,-N/2+2,\ldots,N/2-1,N/2$,
corresponding to the total momentum of the symmetrized state,
and the dimension $d_q$ of the representation $q$ is $1$.
The ground state of the spin-1/2 Heisenberg ring 
for $N=16$ is at the $q=0$ sector and spin singlet.

Figure~\ref{fig.fidelity} shows the fidelity $F\equiv\left|\langle\Psi_0|\Psi^{(q)}(\bm{\theta}_k)\rangle\right|^2$ 
of the ground state between the exact ground state $|\Psi_0\rangle$, calculated with 
the Lanczos exact diagonalization method~\cite{Lin1990,Dagotto1994,Weisse}, 
and the approximate ground state 
$|\Psi^{(q)}(\bm{\theta}_k)\rangle$ obtained after the $k$th iteration of optimizing the variational parameters in the circuit with 
different layer depths $D$. For comparison, the results for the cases with the same circuit structure but not symmetrized 
are also shown. The fidelity $F$ for both symmetrized and non-symmetrized cases is less than $1\%$ when $k=1$ and 
rapidly increases at $k\approx10$. However, the fidelity $F$ is significantly worse
for the non-symmetrized cases, even when 
$D=4$, corresponding to the circuit with $N\times D=64$ variational parameters. In sharp contrast, when the symmetry is 
imposed, the fidelity $F$ becomes as large as $98.8\%$ already for the shallowest circuit with $D=1$ 
and $99.9\%$ with $D=2$, 
clearly demonstrating an excellent improvement by symmetrizing the state.   

Figure~\ref{fig.energy} shows the variational energy of the ground state calculated using 
$|\Psi^{(q)}(\bm{\theta}_k)\rangle$ for both symmetrized and non-symmetrized cases with different layer depth $D$ 
in the circuit. 
As a reference, the exact ground-state energy $E_0$ calculated with
the Lanczos exact diagonalization method is also shown. 
As expected from the fidelity results in Fig.~\ref{fig.fidelity},
the converged variational energy $E^{(q)}(\bm{\theta}_k)$ 
for the non-symmetrized cases is much larger than the exact value $E_0$ even when $D=4$. 
On the other hand, the symmetrized case can obtain the decently accurate energy already
for $D=1$ because $E^{(q)}(\bm{\theta}_{k=10^3})/JN=-0.4447$.
The variational energy is further improved by increasing the number of layers 
to $D=2$, in which 
$E^{(q)}(\bm{\theta}_{k=10^3})/JN=-0.4461$ is essentially exact.

\subsection{Excitation energy}~\label{S1excitation}
One of the advantages of the symmetry-adapted VQE scheme 
is that it can resolve the quantum numbers of the eigenstates 
simply by using the character $\chi^{(q)}(\hat{T}^n)$
of the desired quantum number $q$.
Here we demonstrate this for the lowest magnetically excited states
by calculating the variational energy 
in the $S=1$ sector at momentum $q$,  
\begin{equation}
  E_{S=1}^{(q)}(\bs{\theta})
  \equiv E[\tilde{\Psi}^{(q)}(\bs{\theta})]
  = 
  \frac{
  \langle
  \tilde{\Psi}(\bs{\theta})
  | \hat{\mathcal{H}} 
  \hat{P}^{(q)} |
  \tilde{\Psi}(\bs{\theta})
  \rangle
  }
  {  \langle
  \tilde{\Psi}(\bs{\theta})
  | \hat{P}^{(q)} |
  \tilde{\Psi}(\bs{\theta})
  \rangle},
  \label{eq:s=1}
\end{equation}
where $|\tilde{\Psi}(\bs{\theta})\rangle =
\hat{\mathcal{U}}(\bs{\theta})|\tilde{\Phi}\rangle$ 
with 
\begin{equation}
  |\tilde{\Phi}\rangle =
  \bigotimes_{i=1}^{N/2-1} |s_{2i-1,2i} \rangle
  |t_{N-1,N}\rangle
\end{equation}
and 
$|t_{ij}\rangle = 
(|0\rangle_i |1\rangle_j
+|1\rangle_i |0\rangle_j)/{\sqrt{2}}$.
Note that $|\tilde{\Phi}\rangle$
has the quantum numbers $S=1$ and $S_z=0$~\cite{Oguchi1989} 
and therefore $|\tilde{\Psi}(\bs{\theta})\rangle$ also preserves these quantum numbers. 
The quantum state $|\tilde{\Psi}(\bs{\theta})\rangle$ can be generated
from the same circuit structure in Fig.~\ref{fig.circuit}
merely by setting the initial state at, for example, the $15$th qubit to
$|0\rangle_{15}$, instead of $|1\rangle_{15}$ (see also Appendix~\ref{AppB}).
Notice also that varying the values of $q$ does not require 
any change in the circuit structure, because momentum $q$ 
enters only in the character $\chi^{(q)}(\hat{T}^n)$
[see Eq.~(\ref{eq.Energy})].
Thus, the circuit structure for 
the excited-state calculation remains
the same as that for the ground-state calculation.

\begin{center}
  \begin{figure}
    \includegraphics[width=1\columnwidth]{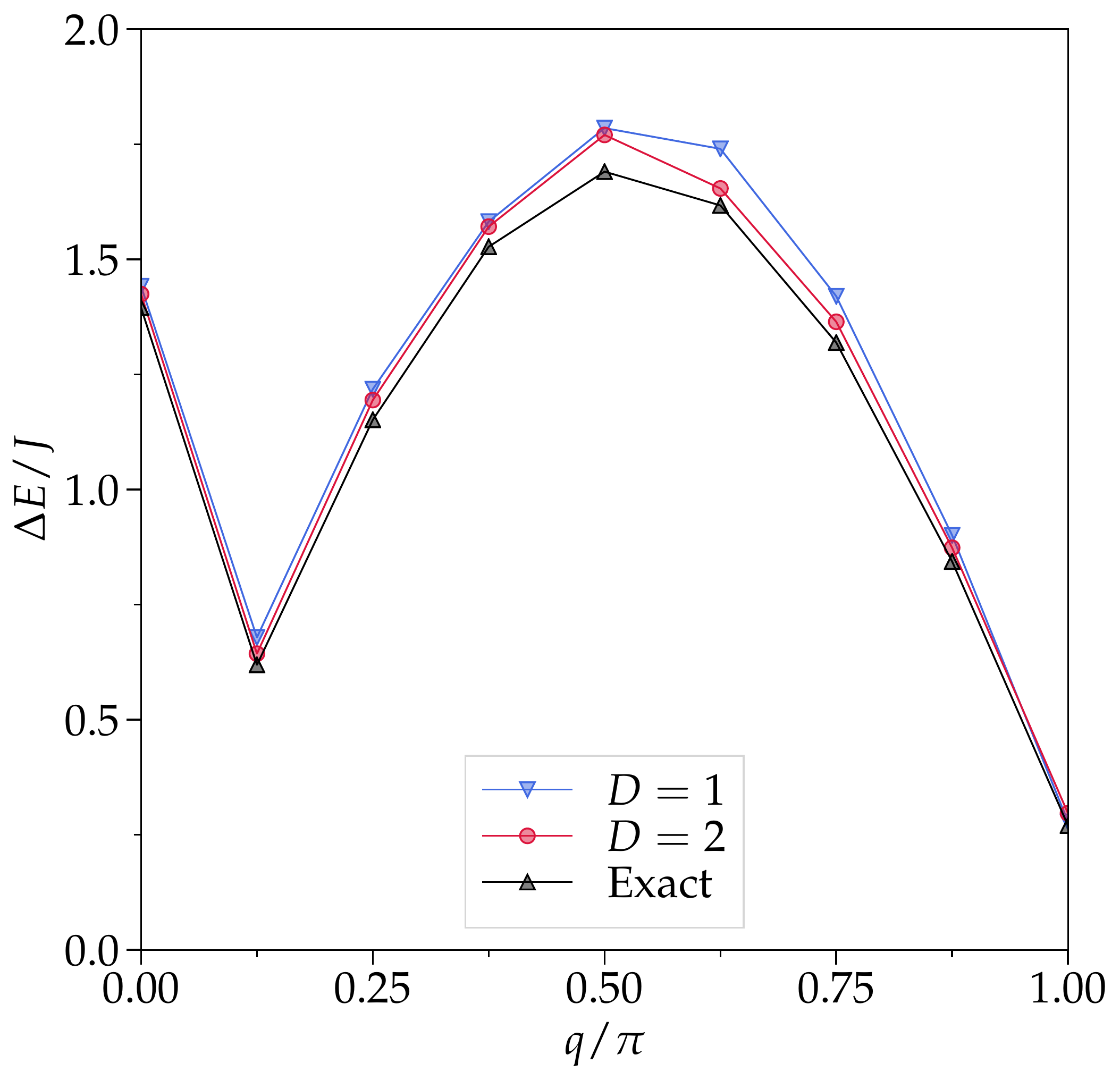}
    \caption{
      Momentum-resolved spin-triplet ($S=1$)
      excitations for the spin-1/2 Heisenberg ring with $N=16$.
      The excitation energy $\Delta E$ is 
      calculated as the difference of the 
      variational energies for the excited state with $S=1$ and momentum $q$ 
      and the ground state. 
      $D$ is the number of layers in the circuit (see Fig.~\ref{fig.circuit}). 
      For comparison, the exact results are also shown. 
      \label{fig.excitation}
    }
  \end{figure}
\end{center}

Figure~\ref{fig.excitation} shows
the spin-triplet excitation energy,
\begin{equation}
  \Delta E \equiv E^{(q)}_{S=1}(\tilde{\bs{\theta}}^*) - E^{(0)}(\bs{\theta}^*), 
\end{equation}
for different momentum $q$, where 
$E^{(0)}(\bs{\theta}^*)$ is the variational energy of the ground state 
discussed in Sec.~\ref{sec:gs} and $E^{(q)}_{S=1}(\tilde{\bs{\theta}}^*) $ 
is the variational energy at the $S=1$ sector with momentum $q$ given in Eq.~(\ref{eq:s=1}). 
$\tilde{\bs{\theta}}^*$ and $\bs{\theta}^*$ are
the optimized variational parameters 
by minimizing separately the corresponding energy functional, 
for which we take the values at the $k=1000$th iteration. 
As shown in Fig.~\ref{fig.excitation}, 
the calculated excitation energies agree well with the exact results already for the shallowest circuit with $D=1$. 
Moreover, 
with increasing the number of layers to $D=2$, 
the accuracy improves systematically, as in the ground-state-energy calculations. 
These results demonstrate that the symmetry-adapted
VQE scheme can also be used to approximate 
low-lying excited states.

\section{Conclusions and discussions}\label{sec.conclusions}
We have proposed a scheme to adapt the Hamiltonian symmetry in the hybrid 
quantum-classical VQE approach. The proposed scheme is to make use of the 
projection operator $\hat{P}^{(\gamma)}_l$ to project a quantum state,
which is described by a quantum circuit that usually breaks the Hamiltonian symmetry
in the VQE approach, 
onto the $l$th basis
of the irreducible representation $\gamma$ of the Hamiltonian symmetry group $\mathcal{G}$.
In the symmetry-adapted VQE scheme proposed here, the nonunitarity 
of the projection operator is treated as postprocessing on classical computers. 
We have also introduced the ``Amida lottery" construction to implement general 
symmetry operations in quantum circuits. Here, each symmetry operation $\hat g$ is simply 
represented as a different product of $O(N)$ SWAP operations and therefore $|\mathcal{G}|$
different circuits are required in the symmetry-adapted VQE scheme. 

Although the symmetry-adapted VQE scheme introduced here is probably the simplest and 
most direct way to implement the Hamiltonian symmetry in the VQE framework, our numerical 
simulations for the spin-1/2 Heisenberg ring clearly demonstrated that the improvement is significant 
in terms of both the fidelity of the ground state and the ground-state energy by showing that the circuit 
with the shallowest layer already achieves the decent accuracy. 
Moreover, we have demonstrated that the symmetry-adapted VQE scheme,
combined with the spin-quantum-number-projected
circuit state, allows us to compute,
for example, the spin-triplet excitation energies
as a function of momentum. 

Recently, a VQE approach with
a Jastrow-type operator, which 
is an exponential of a Hermitian operator   
and is nonunitary in general, 
has been implemented using a quantum hardware~\cite{Mazzola2019}.
While the symmetry projection operator $\hat{P}^{(\gamma)}_l$
is also Hermitian and nonunitary, it is much simpler than 
the Jastrow-type operator,  
in the sense that $\hat{P}^{(\gamma)}_l$
is idempotent and composed of the finite number $|\mathcal{G}|$ of unitary operators. 
In addition, $\hat{P}^{(\gamma)}_l$ commutes with $\hat{\mathcal{H}}$, which  
simplifies the evaluation of
the variational energy, as in Eq.~(\ref{eq.Energy}), and its derivative
with respect to a variational parameter.
We thus expect that the symmetry-adapted VQE approach
described here can be implemented soon with a quantum hardware (also see Appendix~\ref{AppD}).
To this end, an efficient experimental implementation of
SWAP operations 
is highly desirable to perform symmetry operations.

\acknowledgments
We are grateful to
Sandro Sorella for insightful discussion, 
Guglielmo Mazzola for valuable input on quantum computers, and
Yuichi Otsuka for valuable comments.
We are also thankful to RIKEN iTHEMS for providing opportunities for stimulating discussion.   
Parts of numerical simulations have been done
on the HOKUSAI supercomputer at RIKEN 
(Project IDs: G19011 and G20015).
This work was supported by Grant-in-Aid for Research Activity start-up (No.~JP19K23433) and 
Grant-in-Aid for Scientific Research (B) (No.~JP18H01183) from MEXT, Japan, and also by JST PRESTO (No.~JPMJPR191B), Japan.

\appendix

\section{Decomposition of eSWAP gate}~\label{AppA}
\begin{figure}
  \begin{center}
    \includegraphics[width=0.85\columnwidth]{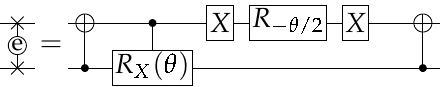}
    \caption{
      A decomposition of the eSWAP gate that is parametrized with $\theta$. 
      \label{circuit_eSWAP}}
  \end{center}
\end{figure}

A decomposition of the eSWAP gate to elementary gates is 
given in Fig.~\ref{circuit_eSWAP}.
Here,
$\hat{R}_{X}(\theta)=\exp(-\imag \theta \hat{X}/2)$ 
and 
$R_{-\theta/2}$ is the phase-shift gate
that acts on a qubit as
$\hat{R}_{-\theta/2}|0\rangle_i = |0\rangle_i$ and
$\hat{R}_{-\theta/2}|1\rangle_i = \e^{-\imag \theta/2}|1\rangle_i$. 
The decomposition in Fig.~\ref{circuit_eSWAP}
can be confirmed readily in the matrix representation as 
\begin{eqnarray}
&&
\begin{bmatrix}
\e^{-\imag \theta/2} & 0 & 0 & 0 \\
0 & \cos\frac{\theta}{2} & -\imag \sin{\frac{\theta}{2}} & 0  \\
0 & -\imag\sin\frac{\theta}{2} & \cos{\frac{\theta}{2}} & 0  \\
0 & 0 & 0 & \e^{-\imag \theta/2}\\
\end{bmatrix}\notag
\\&=& 
\begin{bmatrix}
1 & 0 & 0 & 0 \\
0 & 0 & 0 & 1 \\
0 & 0 & 1 & 0 \\
0 & 1 & 0 & 0 \\
\end{bmatrix}
\begin{bmatrix}
0 & 0 & 1 & 0 \\
0 & 0 & 0 & 1 \\
1 & 0 & 0 & 0 \\
0 & 1 & 0 & 0 \\
\end{bmatrix}
\begin{bmatrix}
1 & 0 & 0 & 0 \\
0 & 1 & 0 & 0 \\
0 & 0 & \e^{- \imag \theta/2}  \\
0 & 0 & 0 & \e^{- \imag \theta/2}  \\
\end{bmatrix}
\begin{bmatrix}
0 & 0 & 1 & 0 \\
0 & 0 & 0 & 1 \\
1 & 0 & 0 & 0 \\
0 & 1 & 0 & 0 \\
\end{bmatrix}\notag 
\\
&\times&
\begin{bmatrix}
1 & 0 & 0 & 0 \\
0 & 1 & 0 & 0 \\
0 & 0 & \cos\frac{\theta}{2} & -\imag \sin{\frac{\theta}{2}}  \\
0 & 0 & -\imag\sin\frac{\theta}{2} & \cos{\frac{\theta}{2}} \\
\end{bmatrix}
\begin{bmatrix}
1 & 0 & 0 & 0 \\
0 & 0 & 0 & 1 \\
0 & 0 & 1 & 0 \\
0 & 1 & 0 & 0 \\
\end{bmatrix},
\label{eq.matrix}
\end{eqnarray}
where the matrix in the left-hand side
represents the eSWAP gate itself [see Eq.~(\ref{eSWAP})], 
and the matrices in the right-hand side represent 
controlled-NOT (CNOT),
controlled-$R_X$,
$X \otimes I$,
$R_{-\theta/2} \otimes I$,
$X \otimes I$, and
CNOT gates, respectively, from right to left in Eq.~(\ref{eq.matrix}). 
Here, the matrices are represented
with respect to the conventional two-qubit basis states 
$|0\rangle_i |0\rangle_j$,
$|0\rangle_i |1\rangle_j$,
$|1\rangle_i |0\rangle_j$, and 
$|1\rangle_i |1\rangle_j$.  
If necessary, the controlled-$R_X$ gate 
can be further decomposed 
into elementary gates~\cite{Barenco1995}. 
From the matrix representation on
the left-hand side of Eq.~(\ref{eq.matrix}), 
it is obvious that the eSWAP gate 
is equivalent to the ${\rm SWAP}^{\alpha}$ gate 
up to a phase factor~\cite{Fan2005,Balakrishnan2008}.

\section{RVB-type state on a quantum circuit}~\label{AppB}

For a physical interpretation of
$|\Psi(\bs{\theta}) \rangle = \hat{\mathcal{U}}(\bs{\theta})|\Phi\rangle$ (see Fig.~\ref{fig.circuit}),
it is important to understand how the SWAP and eSWAP gates act
on the singlet-pair product state $|\Phi\rangle$. 
First, it should be noticed that 
${\hat{\mathcal{P}}}_{ij}$ alters the sign of the wavefunction if it is
operated on the singlet state $|s_{ij}\rangle$ formed between qubits $i$ and $j$:
\begin{equation}
  {\hat{\mathcal{P}}}_{ij}|s_{ij}\rangle=|s_{ji}\rangle = -|s_{ij}\rangle.
  \label{antisym}
\end{equation}
This is simply because the singlet state is antisymmetric with respect
to the permutation of $i$ and $j$. 
In other words, $|s_{ij} \rangle$ is an eigenstate of ${\hat{\mathcal{P}}}_{ij}$
with eigenvalue $-1$. 
The corresponding eSWAP operation results in 
\begin{equation}
  {\hat{U}_{ij}(\theta)}|s_{ij}\rangle= \e^{\imag \theta/2} |s_{ij}\rangle.  
  \label{multphase}
\end{equation}
Thus, $|s_{ij} \rangle$ is an eigenstate of ${\hat{U}_{ij}(\theta)}$ and
operating ${\hat{U}_{ij}(\theta)}$ is equivalent to multiplying a phase factor
on $|s_{ij}\rangle$. 

\begin{figure}
  \begin{center}
    \includegraphics[width=0.95\columnwidth]{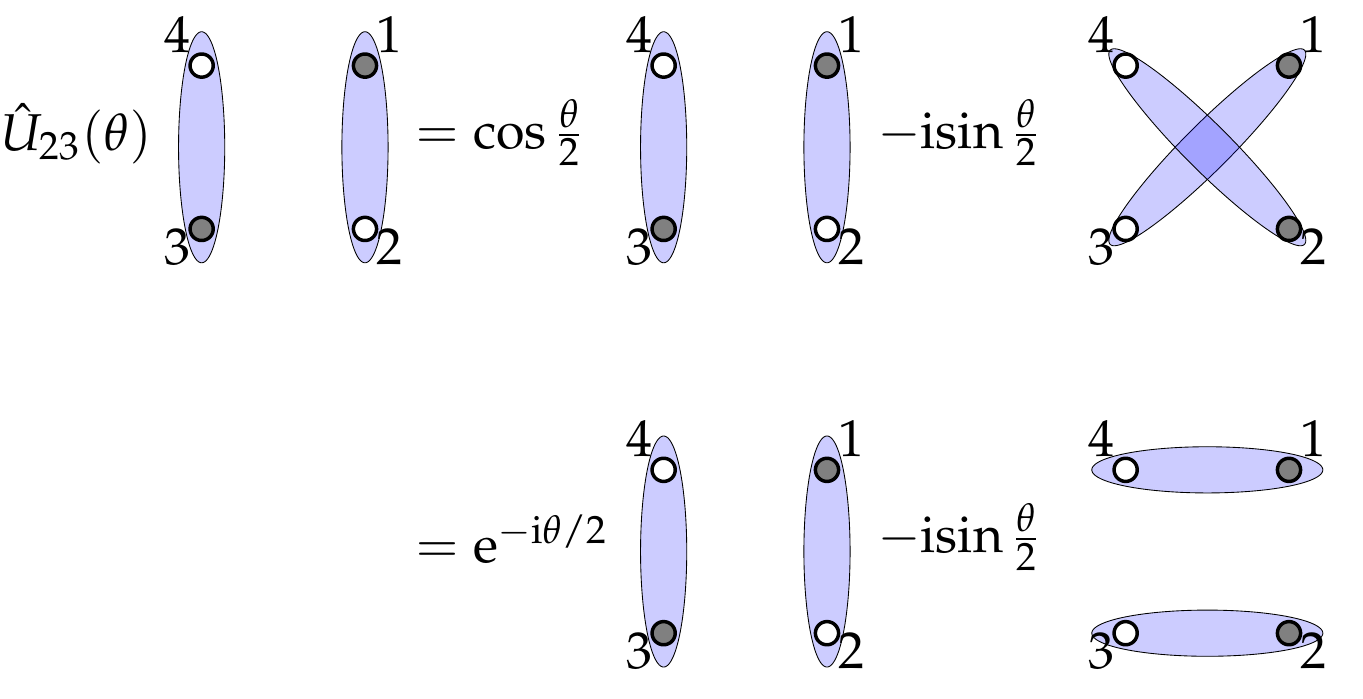}
    \caption{
      A schematic figure of the eSWAP operation
      on a four qubit system.
      An ellipse enclosing two circles (solid and open circles)
      represents a singlet-pair state  
      with the sign convention that, for example, the singlet-pair state formed by qubits 1 and 2 in the left most side, 
      indicated by solid and open circles, respectively, 
      is $|s_{1,2}\rangle =(|0\rangle_1 |1\rangle_2-|1\rangle_1 |0\rangle_2)/\sqrt{2}$. 
      The eSWAP operation between the qubits 2 and 3
      results in a superposition of the different
      singlet-pair product states, i.e., an RVB state. 
      \label{fig.eSWAP23}}
  \end{center}
\end{figure}

If a SWAP gate is operated between two qubits, each of them contributing separately to form different singlets,
then it recombines the singlet pairs as
\begin{equation}
  {\hat{\mathcal{P}}}_{jk} |s_{ij} \rangle |s_{kl} \rangle = |s_{ik} \rangle |s_{jl}\rangle.
  \label{swapsinglet}
\end{equation}
Note that the resulting singlet pairs
are not necessarily formed between the adjacent qubits 
(see for example Refs.~\cite{Kivelson1987,Affleck1988,Tasaki1990}).  
The corresponding
eSWAP operation results in 
\begin{equation}
  {\hat{U}_{jk}(\theta)} |s_{ij} \rangle |s_{kl}\rangle =
          \cos{\frac{\theta}{2}} |s_{ij} \rangle |s_{kl}\rangle
          - \imag \sin{\frac{\theta}{2}} |s_{ik} \rangle |s_{jl}\rangle.
          \label{superpose}
\end{equation}
A crucial feature of the eSWAP gate is that
it not only recombines two singlet pairs 
but also superposes two singlet-pair product states
with parametrized amplitudes.
Namely, the resulting state is a superposition of the original singlet pairs and those generated by the SWAP operation, 
which is essential to generate an RVB state from the  
reference singlet-pair product state $|\Phi \rangle$, 
as will be discussed below. 
Indeed, Eq.~(\ref{superpose}) can already explain how 
an RVB state can be generated on a four qubit system 
(see Fig.~\ref{fig.eSWAP23}).
Notice that the
state represented by the crossed diagram
such as the one in Fig.~\ref{fig.eSWAP23} can be
expressed as a linear combination of those represented by non-crossed
diagrams~\cite{Saito1990}.

The reference state
$|\Phi \rangle$ used here is a dimerized state 
where the singlet pairs are located on the links between adjacent qubits
$(1,2),\, (3,4),\, \ldots,\, (N-1,N)$. 
Such a state breaks the translational symmetry. 
A repeated application of the eSWAP gates, implemented in $\hat{\mathcal{U}}(\bs{\theta})$, 
on $|\Phi\rangle$ 
generates a large number of different
dimer coverings
(configurations of spin-singlet pairs covering all qubits) 
$\mathcal{C}[\Psi(\bs{\theta})]$,
composed of both short-range and long-range singlet pairs~\cite{LR-RVB}, which are 
superposed in the circuit with coefficients parametrized by $\bs{\theta}$.
Thus $|\Psi(\bs{\theta})\rangle$ might be able to
restore the translational symmetry that is broken in $|\Phi \rangle$,
if the number $D$ of layers is large enough.
The present symmetry-adapted VQE scheme,  
instead, restores the spatial symmetry by applying
the projection operator on $|\Psi(\bs{\theta}) \rangle$.

The trial state generated by the circuit that is used in the present study 
thus has a form 
\begin{equation}
  |\Psi (\bs{\theta})\rangle 
  = \sum_{
    \mathcal{C}[\Psi(\bs{\theta})]}
  w(\mathcal{C}[\Psi(\bs{\theta})])
  \bigotimes_{[i,j]\in
    \mathcal{C}[\Psi(\bs{\theta})]}
  |s_{ij} \rangle,
  \label{RVBlike}
\end{equation}
where
$[i,j]$ denotes a pair of two qubits that form $|s_{ij}\rangle$, 
${\mathcal{C}[\Psi(\bs{\theta})]}$ indicates all possible
dimer coverings 
generated on a given circuit, and 
$w(\mathcal{C}[\Psi(\bs{\theta})])$ is a coefficient 
for a singlet-pair product state specified by a configuration ${\mathcal{C}[\Psi(\bs{\theta})]}$. 
It is now obvious that this state in Eq.~(\ref{RVBlike}) has a form of 
the RVB state 
\begin{equation}
  |{\rm RVB} \rangle = \sum_{\mathcal{C}} w(\mathcal{C}) \bigotimes_{[i,j]\in\mathcal{C}} |s_{ij} \rangle, 
\end{equation}
where
${\mathcal{C}}$ denotes all possible
dimer coverings and $w(\mathcal{C})$ is the corresponding coefficient. 
For example, if 
$w(\mathcal{C})$ is taken to be equally weighted
for all the configurations that consist of only nearest-neighbor singlet pairs, 
$|{\rm RVB}\rangle$ reduces to a so-called short-range RVB state
(see for example Ref.~\cite{Yunoki2006} for a detailed description). 
However, we should emphasize the important difference between $|\Psi (\bs{\theta})\rangle$ 
and $|{\rm RVB} \rangle$. While all the coefficients $w(\mathcal{C})$ in $|{\rm RVB} \rangle$ can be set independently 
for different realizations of all possible dimer coverings ${\mathcal{C}}$, the coefficients 
$w(\mathcal{C}[\Psi(\bs{\theta})])$ in $|\Psi (\bs{\theta})\rangle$ are not independent
but related to each other via the variational 
parameters $\bm{\theta}$ in the circuit even though the repeated application of the eSWAP gates 
can eventually produce all possible dimer coverings.
The RVB state 
has often been used as a variational wavefunction
for approximating the ground states of
the spin-1/2 Heisenberg model in square~\cite{Liang1988,Poilblanc1989},
triangular~\cite{Oguchi1986}, and kagome lattices~\cite{Sindzingre1994}. 
A numerical study on small clusters
up to 26 spins~\cite{Nakagawa1990} has shown
that, by taking into account the Marshall's sign rule~\cite{Marshall1955},
the RVB state with only a few variational parameters
can accurately represent the ground state
of the spin-1/2 Heisenberg model in a square lattice,
and that the (long-range) RVB state
substantially improves the variational energy and the variational state 
as compared to the short-range RVB state. 

Finally, we briefly note 
on calculations in higher spin-quantum-number sectors 
assuming that $N$ is even. 
One can derive relations similar to Eqs.~(\ref{antisym})--(\ref{superpose}) 
for the spin-triplet states 
$|t_{ij} \rangle \equiv (|0\rangle_{i}|1\rangle_{j} + |1\rangle_i |0\rangle_j)/\sqrt{2}$,
$|t_{ij}^+ \rangle \equiv |0\rangle_i |0\rangle_j$, and 
$|t_{ij}^- \rangle \equiv |1\rangle_i |1\rangle_j$. 
A difference here from the case of $|s_{ij}\rangle$
is that the triplet states are 
symmetric under the SWAP operation. 
By using a product state of
$N/2-1$ singlet pairs and a single triplet pair $|t_{ij} \rangle$,
instead of $|\Phi\rangle$, 
as the reference state, 
one can search for the lowest-energy state within the subspace
of $S=1$ and $S_z=0$, as demonstrated in Sec.~\ref{S1excitation}
(see Ref.~\cite{Oguchi1989} for a detailed analysis).
The calculation in the higher $S$ sectors with
finite-$S_z$ states is also possible simply by using $|t^+_{ij}\rangle$ or $|t^-_{ij}\rangle$ 
for the reference state. 
Finding the lowest energy in the higher spin sectors 
is useful for studying, for example,  whether
a magnetic long-range order exists in the thermodynamic limit from finite-size
calculations~\cite{Bernu1994,Sindzingle2002,Sindzingle2004,Shanon2006}.

Such a circuit explicitly specifies the subspace
labeled by the spin-quantum numbers $S$ and $S_z$,
and thus is specialized to 
spin-isotropic (i.e., SU(2) symmetric) Heisenberg models. 
On classical computers,  
with a sophisticated and elaborated algorithm
that incorporates the spatial symmetry, such as
the lattice translational symmetry, and $S_z$ conservation~\cite{Lin1990,Weisse2013}, 
one can obtain the numerically exact ground state of
the spin-1/2 Heisenberg model up to $50$ spins~\cite{Wietek2018},
which is far larger than the case of $16$ qubits studied here.
However, 
$\bs{S}^2$ conservation is usually not implemented 
because the programming of a total-spin-preserved code is, 
although possible~\cite{Bostrem2006,Heitmann2019}, 
not easy and often computationally demanding
on classical computers. 
We expect that
the circuit that operates eSWAP gates
on a singlet-pair product state or
on a pair-product state with higher spin-quantum numbers 
might be useful for studying spin-liquid states including the RVB state 
as well as excited states 
on quantum computers in the near future.
Regarding excitations and dynamics,
we should also note that the eSWAP operations 
naturally appear also in such simulations 
when a Suzuki-Trotter decomposition is applied to
the time-evolution operator $\e^{-\imag \hat{\mathcal{H}} t}$
with $t$ being time~\cite{Suzuki1976,Trotter1959}. 

\section{Generation of spin-singlet pairs formed by distant qubits}\label{AppC}
In this Appendix, we show that
a spin-singlet pair formed by qubits that are separated at the largest distance 
can still be generated by repeated application of the nearest-neighbor eSWAP gates on the
singlet-pair product state $|\Phi\rangle$ with $D~\sim N/4$ for 
the one-dimensional chain of $N$ qubits under the periodic boundary conditions.

\begin{figure}
  \begin{center}
    \includegraphics[width=0.95\columnwidth]{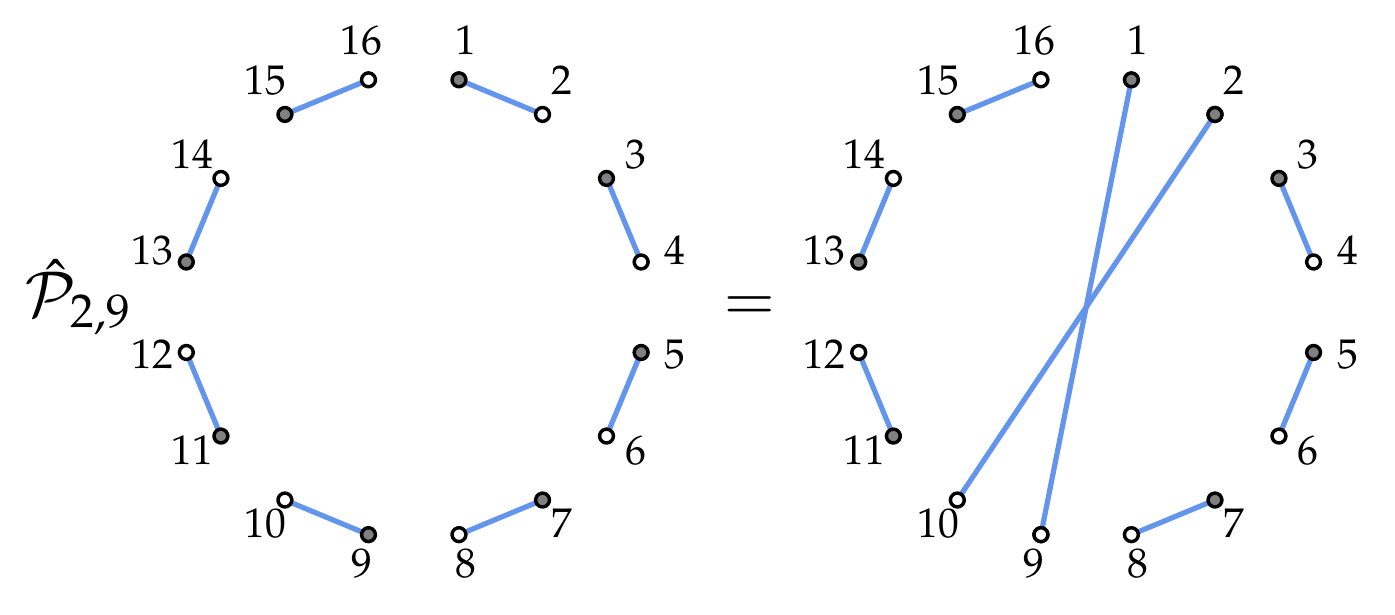}\\
    \caption{
      Schematic figure of a long-range SWAP operation 
      on the singlet-pair product state $|\Phi\rangle$ 
      for $N=16$ and $r=8$.
      Here a singlet-pair state is represented by a
      blue line ending with solid and open circles.
      \label{longSWAP}
      }
    \end{center}
\end{figure}

Let us first consider
how a spin-singlet state with $r$-lattice spacing, e.g., 
\begin{equation}
  |s_{1,1+r} \rangle, 
\end{equation}
can be generated from the singlet-pair product state $|\Phi\rangle$. 
Here, to be specific, we assume that $N$ and $r$ are both even.  
According to Eq.~(\ref{swapsinglet}), 
a long-range SWAP operation $\hat{\mathcal{P}}_{2,r+1}$ on two
nearest-neighbor spin-singlet pairs 
$|s_{1,2}\rangle$ and $|s_{r+1,r+2}\rangle$ generates such $r$-distant singlet pairs:
\begin{equation}
  \mathcal{\hat{P}}_{2,r+1} |s_{1,2}\rangle |s_{r+1,r+2}\rangle
  = |s_{1,r+1}\rangle |s_{2,r+2}\rangle.
\end{equation}
Figure~\ref{longSWAP} illustrates 
the generation of spin-singlet pairs formed by distant qubits in the case of
$N=16$ and $r=8$.

\begin{figure*}
  \begin{center}
    \includegraphics[width=1.95\columnwidth]{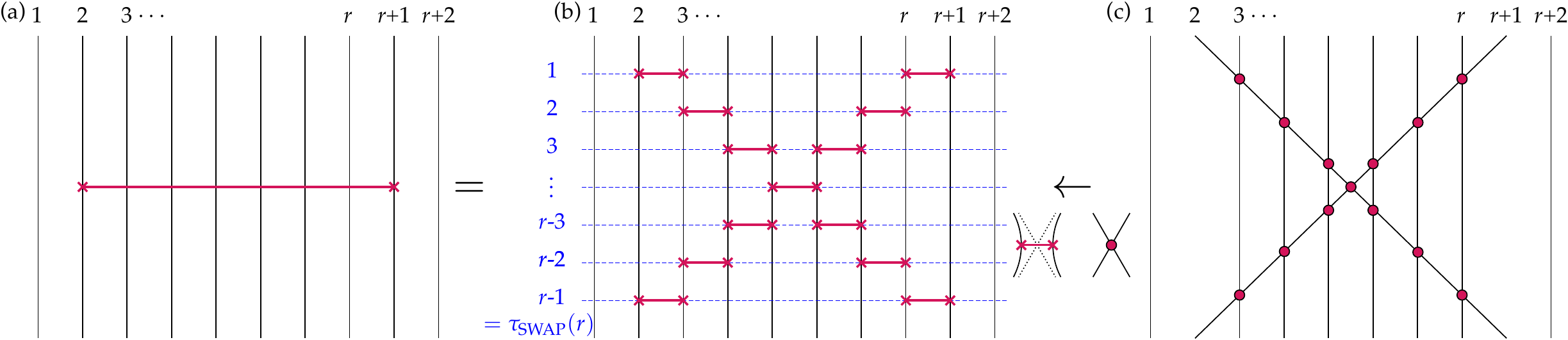}\\
    \caption{
      A decomposition of a long-range SWAP gate with the 
      ``Amida lottery'' construction. 
      (a) The long-range SWAP gate $\hat{\mathcal{P}}_{2,r+1}$ with $r=8$.
      (b) A decomposition of $\hat{\mathcal{P}}_{2,r+1}$ to
      the nearest-neighbor SWAP gates. 
      The horizontal dashed lines associated with the numbers  
      $1,2, \cdots, \tau_{\rm SWAP}(r)=r-1$ indicate the time steps. 
      (c) Auxiliary figure that generates the decomposition of
      the long-range SWAP gate into the nearest-neighbor SWAP gates shown in (b).
      \label{longSWAPamida}
      }
    \end{center}
\end{figure*}

Next, we consider how the long-range SWAP operator 
$\hat{\mathcal{P}}_{2,r+1}$ can be represented as a product of the
nearest-neighbor SWAP operators.
For this purpose, we make use of the ``Amida lottery'' construction introduced in Sec.~\ref{sec.symop}.
Figure~\ref{longSWAPamida} shows that, following the ``Amida lottery'' construction, the long-range SWAP operation
can indeed be expressed as a product of the nearest-neighbor SWAP
operations that form an X-like shape on the circuit. 
The number $N_{\rm SWAP}(r)$ of the nearest-neighbor SWAP gates necessary in the circuit 
is 
\begin{equation}
  N_{\rm SWAP}(r) = 2(r-2)+1 = 2r-3, 
\end{equation}
as there are $r-2$ qubits between the $2$nd and $(r+1)$st qubits (see Fig.~\ref{longSWAPamida}). 
One can also find that, with this construction, 
the depth of the circuit or  the number of ``time steps'' $\tau_{\rm SWAP}(r)$ required is
\begin{equation}
  \tau_{\rm SWAP}(r) = (r-2)+1=r-1. 
\end{equation}

\begin{figure*}
  \begin{center}
    \includegraphics[width=1.95\columnwidth]{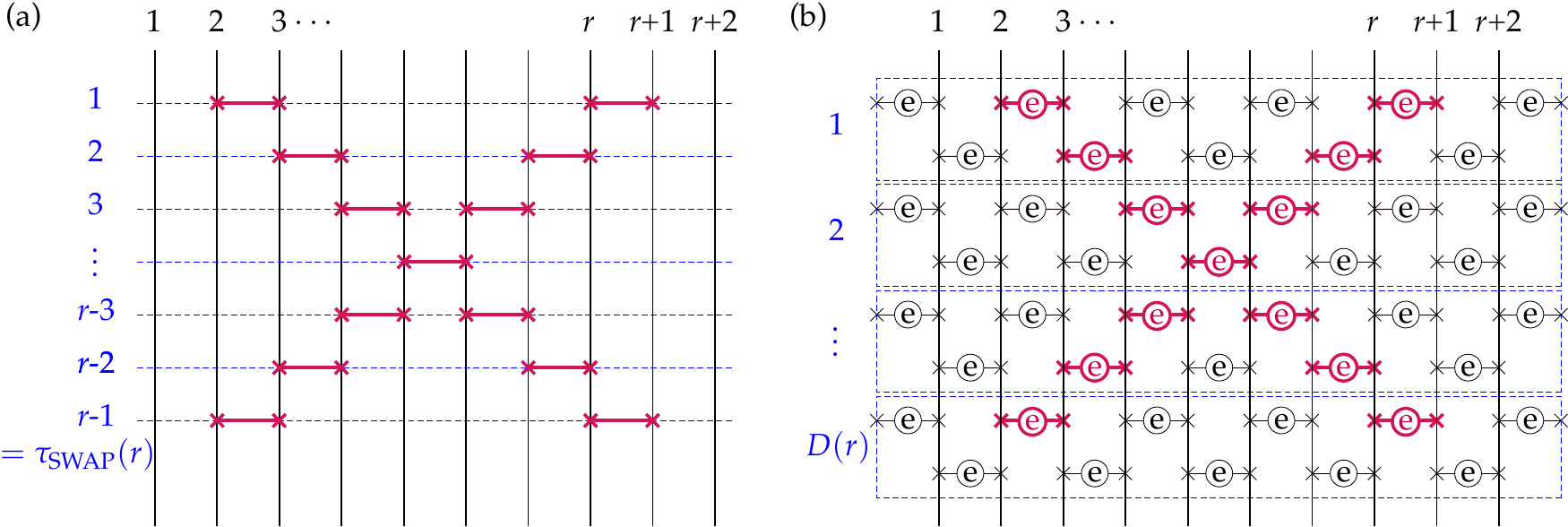}\\
    \caption{
      (a) The long-range SWAP gate $\hat{\mathcal{P}}_{2,r+1}$ with $r=8$ generated by a set of the nearest-neighbor SWAP gates 
      (see Fig.~\ref{longSWAPamida}).  
      (b) An equivalent operation (up to a global phase factor) can be described by 
      the sequence $\hat{\mathcal{U}}(\bs{\theta}_{\rm SWAP})$ of the nearest-neighbor eSWAP gates, where 
      the rotation angles $\theta_{ij}$ are either $\pi$ or $-\pi$
      for the eSWAP gates highlighted with red thick lines, 
      and $\theta_{ij}=0$, otherwise.
      Each dashed box associated with the number $1,2,\cdots$, or $D(r)$ in (b) corresponds to the single layer of the parametrized gates 
      indicated by shaded blue in Fig.~\ref{fig.circuit}.       
      } \label{longSWAPeSWAP}
    \end{center}
\end{figure*}

Noticing that 
$\hat{U}_{ij}(0)=\hat{I}$ and
$\hat{U}_{ij}(\pm\pi)=\mp\imag \hat{\mathcal{P}}_{ij}$ in Eqs.~(\ref{eSWAP}) and (\ref{eq:U}), 
we can now readily show that
the sequence $\hat{\mathcal{U}}(\bs{\theta})$
of the nearest-neighbor eSWAP operations in Fig.~\ref{fig.circuit} 
with a particular set of parameters $\bs{\theta}=\bs{\theta}_{\rm SWAP}$
can produce $\hat{\mathcal{P}}_{2,r+1}$,
up to a global phase factor, i.e.,
\begin{equation}
  \hat{\mathcal{U}}(\bs{\theta}_{\rm SWAP}) = \pm \imag \hat{\mathcal{P}}_{2,r+1}. 
  \label{UPr}
\end{equation}
Namely, $\bs{\theta}_{\rm SWAP}$ has $\theta_{ij}=\pm \pi$
if $\langle i,j \rangle$ corresponds to the link
on which the nearest-neighbor SWAP operation
is required for $\hat{\mathcal{P}}_{2,r+1}$,
and $\theta_{ij}=0$, otherwise (see Fig.~\ref{longSWAPeSWAP}).
The global phase factor, which is however irrelevant
for the purpose of this Appendix, in Eq.~(\ref{UPr}) appears because
$N_{\rm SWAP}(r)$ is odd, and depends on how the sign
of $\theta_{ij}=\pm \pi$ is chosen. 
The number $D(r)$ of layers in $\hat{\mathcal{U}}(\bs{\theta}_{\rm SWAP})$
required for producing $\hat{\mathcal{P}}_{2,r+1}$ is thus  
\begin{equation}
  D(r)=\left \lceil \frac{\tau_{\rm SWAP}(r)}{2} \right\rceil
  =\left \lceil \frac{r}{2}-\frac{1}{2} \right\rceil,
  \label{Depth_r}
\end{equation} 
where $\lceil \cdot \rceil$ denotes the ceiling function which 
returns the minimum integer larger than or equal to the argument.
The argument in Eq.~(\ref{Depth_r}) is divided
by $2$ because each layer of 
$\hat{\mathcal{U}}(\bs{\theta})$ contains 
two time steps (see Fig.~\ref{longSWAPeSWAP}).

Under the periodic-boundary condition,
the largest distance $r_{\rm max}$ is 
\begin{equation}
  r_{\rm max}=N/2. 
\end{equation}
Therefore, to generate 
a spin-singlet pair formed by qubits separated at the largest distance, 
the required number of layers is  
\begin{equation}
  D(r_{\rm max})
  =\left \lceil \frac{N}{4}-\frac{1}{2} \right\rceil,
  \label{Depth_rmax} 
\end{equation} 
i.e., $D(r_{\rm max})\sim N/4$. 
However, this does {\it not} necessarily imply that 
all possible dimer coverings are generated with $D=D(r_{\rm max})$.  

Finally, we note that 
since $\bs{\theta}$ in general takes
arbitrary values, 
$|\Psi(\bs{\theta})\rangle = \hat{\mathcal{U}}(\bs{\theta})|\Phi\rangle$
is a superposition of many different singlet-product states represented by 
different dimer coverings, 
among which spin-singlet pairs formed by distant qubits are certainly contained, 
as discussed above, although only the nearest-neighbor eSWAP gates are applied in the circuit.

\section{Simulation on quantum hardware}\label{AppD}
To validate the relevance of the RVB-type wavefunction
as a trial wavefunction on quantum computers, 
in this Appendix we estimate the ground state energy 
for a small system ($N=4$) using the ibmqx2 chip,
which consists of five qubits, available through 
an online quantum computing network provided by IBM (IBM Q 5 Yorktown)~\cite{IBM} 
with the Qiskit \textsc{python} API for programming the device~\cite{Qiskit}.

Let us first review the ground-state properties of 
the spin-$1/2$ Heisenberg model on the $N=4$ ring. 
With the labeling of qubits shown in Fig.~\ref{fig.eSWAP23}, 
the exact ground state $|\Psi_0\rangle$ is given by
\begin{equation}
  |\Psi_0\rangle =
  \frac{1}{\sqrt{3}}
  \left(
  |s_{1,2}\rangle |s_{3,4}\rangle
  +
  |s_{4,1}\rangle |s_{2,3}\rangle
  \right). 
\end{equation}
$|\Psi_0\rangle$ is a superposition
of the two singlet-pair product states
with the same probability amplitude, and 
is correctly normalized 
because these singlet-pair product states are not
orthogonal to each other but have an overlap of 
$(\langle s_{4,1}| \langle s_{2,3}|)(|s_{1,2}\rangle |s_{3,4}\rangle)=1/2$. 
The corresponding exact ground-state energy is 
\begin{equation}
  E_0=-2J.
  \label{eq.E0exact}
\end{equation}
In terms of the expectation value of the Hamiltonian,
$E_0$ is expressed as 
\begin{equation}
  E_0
  = \frac{J}{4} 
  \sum_{i=1}^{N}
  \langle \Psi_0 |\left(
   \hat{X}_i \hat{X}_{i+1}
  +\hat{Y}_i \hat{Y}_{i+1}
  +\hat{Z}_i \hat{Z}_{i+1}
  \right) |\Psi_0 \rangle, 
\end{equation}
where $i+1$ should be identified as $1$ if $i=N$
because of the periodic-boundary conditions. 
Since $|\Psi_0\rangle$ is spin-symmetric
and translationally invariant, 
Eq.~(\ref{eq.E0exact}) can be rephrased in terms of 
the exact nearest-neighbor spin correlation functions as 
\begin{alignat}{1}
  &\langle \Psi_0 | \hat{X}_i \hat{X}_{i+1} | \Psi_0 \rangle
  =\langle \Psi_0 | \hat{Y}_i \hat{Y}_{i+1} | \Psi_0 \rangle
  =\langle \Psi_0 | \hat{Z}_i \hat{Z}_{i+1} | \Psi_0 \rangle \notag \\
  =&
  \frac{E_0}{3N(J/4)}
  = -\frac{2}{3}
  \label{eq.XXYYZZ}
\end{alignat}
for any $i$.

Next we show that, up to a global phase factor, $|\Psi_0\rangle$ 
can be produced by 
applying two eSWAP gates on the singlet-pair product 
state $|s_{1,2}\rangle|s_{3,4}\rangle$. 
A straightforward calculation with 
Eqs.~(\ref{multphase}) and (\ref{superpose}) shows that 
\begin{equation}
  \hat{U}_{34}(\theta_2) \hat{U}_{23}(\theta_1) |s_{1,2}\rangle|s_{3,4}\rangle=  \e^{\imag \phi}|\Psi_0\rangle,  
\end{equation}
where 
\begin{alignat}{1}
  &{\theta_1} = 2 \arccos{\left(-\sqrt{\frac{2}{3}}\right)} = 1.6081734479693928 \times \pi, \\
  &{\theta_2} = 2 \arccos{\left(-\sqrt{\frac{1}{3}}\right)} = 1.3918265520306072 \times \pi,  
\end{alignat}
and $\e^{\imag \phi}=\sqrt{\frac{2}{3}}-\sqrt{\frac{1}{3}}\imag$.
Hereafter, we ignore the global phase factor $\e^{\imag \phi}$ because
it is irrelevant for the energy estimation.

\begin{figure*}
\begin{center}
  \includegraphics[width=1.95\columnwidth]{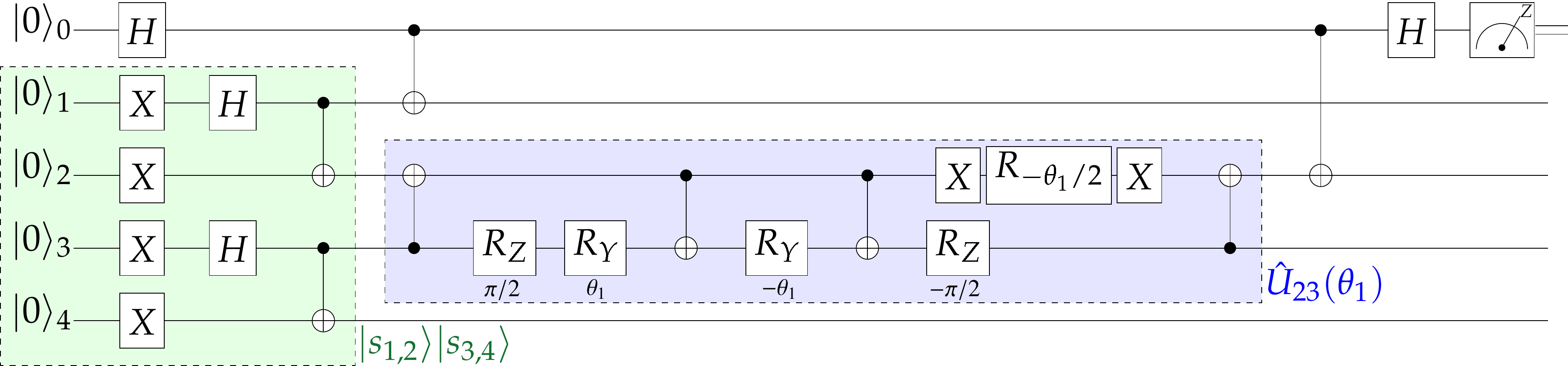}\\
  \caption{    
    The circuit used for evaluating
    ${\rm Re}\langle \Psi_0 |\hat{X}_1 \hat{X}_2 |\Psi_0 \rangle =
    {\rm Re}\langle \tilde{\Psi}_0 |\hat{X}_1 \hat{X}_2 | \tilde{\Psi}_0 \rangle $
    on the ibmqx2 chip. 
    The state 
    $|\tilde{\Psi}_0\rangle=\hat{U}_{23}(\theta_1)|s_{1,2}\rangle |s_{3,4} \rangle$
    is generated on the first to fourth qubits.  
    The parts of the circuit corresponding to 
    $|s_{1,2}\rangle |s_{3,4} \rangle$ and $\hat{U}_{23}(\theta_1)$ are
    highlighted with shaded green and blue boxes, respectively.
    The rotation angles for $R_Y$ and $R_Z$ gates are also indicated below these gates.
    \label{fig.ibmq}
  }
\end{center}
\end{figure*}

Now we consider the energy estimation on quantum computers. 
Equation~(\ref{eq.XXYYZZ}) implies that
evaluating one of these correlation functions  
suffices for estimating $E_0$. 
Here, we evaluate the correlation function
$\langle \Psi_0 | \hat{X}_1 \hat{X}_2 | \Psi_0 \rangle$
by using the Hadamard test as 
\begin{equation}
  {\rm Re}\langle \Psi_0 | \hat{X}_1 \hat{X}_2 | \Psi_0 \rangle
  = p_0 - p_1, 
  \label{eq.ReXX}
\end{equation}
where
\begin{alignat}{1}
  p_0 & = \frac{1}{2}\left(1 + {\rm Re}\langle \Psi_0|\hat{X}_1\hat{X}_2|\Psi_0 \rangle \right) 
\end{alignat}
and
\begin{alignat}{1}
  p_1 & = \frac{1}{2}\left(1 - {\rm Re}\langle \Psi_0|\hat{X}_1\hat{X}_2|\Psi_0 \rangle \right) 
\end{alignat}
are probabilities
of observing $0$ and $1$, respectively, by measuring out
the ancilla ($0$th) qubit in Fig.~\ref{fig.ibmq}~\cite{DM}. 
Among the correlation functions, 
$\hat{X}_1 \hat{X}_2$ is chosen because CNOT gate is 
implemented as one of the basis gates on the ibmqx2 chip.
Moreover,
since $\hat{X}_1 \hat{X}_2$ does not involve qubits $3$ and $4$,
operation of $\hat{U}_{34}(\theta_2)$ is not necessary for 
measurements of $\hat{X}_1 \hat{X}_2$.  
Namely, since
$\left[\hat{X}_1 \hat{X}_2, \hat{U}_{34}(\theta)\right]=0$ for any $\theta$, 
the correlation function can be simplified as 
\begin{alignat}{1}
   \langle \Psi_0 | \hat{X}_1 \hat{X}_2 | \Psi_0 \rangle
   &=\langle \tilde{\Psi}_0 |
   \hat{U}_{34}(\theta_2)^\dag
   \hat{X}_1 \hat{X}_2
   \hat{U}_{34}(\theta_2)
   | \tilde{\Psi}_0 \rangle \notag \\
  & =\langle \tilde{\Psi}_0 | \hat{X}_1 \hat{X}_2 | \tilde{\Psi}_0 \rangle, 
\end{alignat}
where 
\begin{equation}
  |\tilde{\Psi}_0\rangle = \hat{U}_{23}(\theta_1) |s_{1,2}\rangle|s_{3,4}\rangle. 
\end{equation}
On the ibmqx2 chip,
we implement a circuit that generates 
$|\tilde{\Psi}_0\rangle$ 
for measurements. The eSWAP gate
corresponding to $\hat{U}_{23}(\theta_1)$ is implemented
with the decomposition shown in Fig.~\ref{circuit_eSWAP}, 
where the controlled-$R_X$ gate is further decomposed in the way described in Ref.~\cite{Barenco1995}.

Table~\ref{table.P0} shows the probabilities $p_0$ and $p_1$,  
and estimated values of 
${\rm Re}\langle \Psi_0 | \hat{X}_1 \hat{X}_2 | \Psi_0 \rangle$
from 16 samples, each of which consists of
1024 measurements.
The negative values of 
${\rm Re}\langle \Psi_0 | \hat{X}_1 \hat{X}_2 | \Psi_0 \rangle$
imply the antiferromagnetic correlation between
the nearest-neighbor spins.
In the ideal (noiseless) case, the probabilities are 
$p_0=1/6$ and $p_1=5/6$.
Averaging over the results of the 16 samples yields 
${\rm Re}\langle \Psi_0 | \hat{X}_1 \hat{X}_2 | \Psi_0 \rangle = -0.66894(549)$ 
and hence $E_0/J=-2.00682(1647)$, 
where the numbers in parentheses represent 
the standard error of the mean for the last digits.
Therefore, the exact energy is obtained within the statistical error.

\begin{table}
\begin{center}
  \caption{Probabilities $p_0$ and $p_1$ 
    obtained from quantum simulations on the ibmqx2 chip.
    The values on each row are obtained from 1024 measurements.
    Ideal (noiseless) results are also shown in the bottom row. 
    Data were obtained on 6 April 2020 (EST)~\cite{DATA}.
    \label{table.P0}
  }
  \begin{tabular}{
      m{0.24\columnwidth}
      >{\centering\arraybackslash}m{0.24\columnwidth}
      >{\centering\arraybackslash}m{0.24\columnwidth}
      >{\centering\arraybackslash}m{0.24\columnwidth}
    }
  \hline
  \hline
  Sample & $p_0(\%)$ & $p_1(\%)$ & ${\rm Re}\langle \Psi_0| \hat{X}_1 \hat{X}_2 |\Psi_0 \rangle$ \\
  \hline  
  1 & 15.430 & 84.570 & -0.69140 \\
2 & 17.969 & 82.031 & -0.64062 \\ 
3 & 15.625 & 84.375 & -0.68750 \\
4 & 16.309 & 83.691 & -0.67382 \\ 
5 & 16.016 & 83.984 & -0.67968 \\ 
6 & 15.430 & 84.570 & -0.69140 \\ 
7 & 17.578 & 82.422 & -0.64844 \\ 
8 & 18.457 & 81.543 & -0.63086 \\ 
9 & 17.090 & 82.910 & -0.65820 \\ 
10 & 17.969 & 82.031 & -0.64062 \\ 
11 & 16.602 & 83.398 & -0.66796 \\ 
12 & 17.090 & 82.910 & -0.65820 \\ 
13 & 16.992 & 83.008 & -0.66016 \\ 
14 & 15.527 & 84.473 & -0.68946 \\ 
15 & 16.113 & 83.887 & -0.67774 \\ 
16 & 14.648 & 85.352 & -0.70704 \\ 
\hline 
Mean & 16.553(274) & 83.447(274) & -0.66894(549) \\
\hline
Ideal & 16.667 & 83.333 & -0.66667 \\ 
\hline
  \hline
\end{tabular}
\end{center}
\end{table}

It is interesting to note that the ground-state energy obtained here is significantly better than 
the one estimated with the hardware-efficient ansatz reported in Ref.~\cite{Kandala2017},
where the ground-state energy is approximately $-1.5J$~\cite{H_IBM}. 
The substantial improvement found here over the circuit based on the hardware-efficient ansatz  
is highly instructive and suggests 
that the construction of quantum circuits based on the RVB-type wavefunction, which takes into account the spin rotational symmetry, 
is a better strategy to describe the ground state (and also excited states) of
the Heisenberg model on quantum computers.

Finally, we comment on quantum simulations 
of the same system with the symmetry-projection scheme. 
Unfortunately, we have found it difficult 
to implement the symmetry operators on a real quantum device at present. 
The difficulty is due to controlled-SWAP (Fredkin) gates, 
each of which is decomposed into many CNOT gates and one-qubit rotations,  
causing formidably noisy results.
An efficient implementation of the controlled-SWAP (Fredkin) gate in a quantum device, 
as demonstrated in Ref.~\cite{Patel2016}, 
is thus highly desirable.

\bibliography{biball}

\begin{thebibliography}{112}
\bibitem [{PJO()}]{PJO}%
  \BibitemOpen
  \href@noop {} {}\bibinfo {note} {When a matrix representation is not
  available, we can use the character $\chi^{(\gamma)}(\hat{g})=\sum_l
  \left[{\bar D}^{(\gamma)}(\hat{g})\right]_{ll}$ of the irreducible
  representation $\gamma$ for the symmetry operation $\hat{g}$ and introduce
  the following projection operator: \begin{equation} \hat{P}^{(\gamma)} =
  \frac{d_{\gamma}}{|\mathcal{G}|} \sum_{\hat{g}\in\mathcal{G}}
  \chi^{(\gamma)}(\hat{g})^* \hat{g}, \nonumber \end{equation} which projects
  an arbitrary quantum state $|\psi\rangle$ onto a state in the irreducible
  representation $\gamma$, but in general composed of a sum of the $d_\gamma$
  base functions that span the subspace of the irreducible representation
  $\gamma$~\cite{Inui}}\BibitemShut {NoStop}%
\bibitem [{not()}]{note1}%
  \BibitemOpen
  \href@noop {} {}\bibinfo {note} {Here, we assume that the irreducible
  representation $\gamma$ is one-dimensional. However, the extension to the
  case of a higher-dimensional irreducible representation is straightforward by
  using the projection operator $\hat{P}^{(\gamma)}_l$ in
  Eq.~(\ref{eq:pjo})}\BibitemShut {NoStop}%
\bibitem [{AV()}]{AV}%
  \BibitemOpen
  \href@noop {} {}\bibinfo {note} {As long as an operator $\hat{\mathcal O}$
  commutes with $\hat{\mathcal{H}}$, the expectation value of $\hat{\mathcal
  O}$ is evaluated similarly as in Eq.~(\ref{eq.Energy}). However, when this is
  not the case, the numerator of the corresponding formula for the expectation
  value is slightly more complex and the computational complexity increases by
  an additional $\mathcal{O}(N)$ factor}\BibitemShut {NoStop}%
\bibitem [{LR-()}]{LR-RVB}%
  \BibitemOpen
  \href@noop {} {}\bibinfo {note} {This is the case even though the eSWAP gates
  are repeatedly applied only between adjacent qubits, as shown in
  Fig.~\ref{fig.circuit}. More details are discussed in
  Appendix~\ref{AppC}}\BibitemShut {NoStop}%
\bibitem [{DM()}]{DM}%
  \BibitemOpen
  \href@noop {} {}\bibinfo {note} {We can certainly measure directly the
  correlation functions in Eq.~(\ref{eq.XXYYZZ}) without introducing any
  ancilla qubit. However, we find that the accuracy is much worse, for example,
  when we directly measure $\hat Z_1 \hat Z_2$ using 4 qubits in the ibmqx2
  chip, obtaining $\langle \Psi_0|\hat{Z}_1\hat{Z}_2|\Psi_0\rangle=
  -0.41371(845)$ for the same number of measurements as in
  Table~\ref{table.P0}}\BibitemShut {NoStop}%
\bibitem [{DAT()}]{DATA}%
  \BibitemOpen
  \href@noop {} {}\bibinfo {note} {We have obtained essentially the same
  results on different dates and thus under different calibration
  conditions}\BibitemShut {NoStop}%
\end{thebibliography}

\end{document}